\documentclass[aps,prb,twocolumn,showpacs,floatfix,amsmath,amssymb]{revtex4}

\usepackage{amsmath,bm,amsfonts}
\usepackage{graphicx}

\newfont{\bg}{cmr10 scaled\magstep4}

\newcommand{\bigzerou}{\smash{\lower1.7ex\hbox{\bg 0}}}

\begin{document}

\title{Numerically exact O($N^{7/3}$) method 
       for large-scale electronic structure calculations}
  
\author{Taisuke Ozaki}
\address{
    Research Center for Integrated Science (RCIS), 
    Japan Advanced Institute of Science and Technology (JAIST),
    1-1 Asahidai, Nomi, Ishikawa 923-1292, Japan
}

\date{\today}

\begin{abstract} 
An efficient low-order scaling method is presented for large-scale 
electronic structure calculations based on the density functional 
theory using localized basis functions, 
which directly computes selected elements of the density matrix 
by a contour integration of the Green function evaluated 
with a nested dissection approach for resultant sparse matrices. 
The computational effort of the method scales as O($N(\log_2N)^2$), 
O($N^{2}$), and O($N^{7/3}$) for one, two, and three dimensional
systems, respectively, where $N$ is the number of basis functions.
Unlike O($N$) methods developed so far the approach is a numerically 
exact alternative to conventional O($N^{3}$) diagonalization schemes 
in spite of the low-order scaling, and can be applicable to not only 
insulating but also metallic systems in a single framework.
It is also demonstrated that the nested algorithm and the well separated 
data structure are suitable for the massively parallel computation,
which enables us to extend the applicability of density functional 
calculations for large-scale systems together with the low-order scaling.

\end{abstract}


\maketitle

\section{INTRODUCTION}

During the last three decades continuous efforts
\cite{CP,Payne,Davidson,Pulay,Wood,Teter,Stich,Kresse,Householder,Golub,Goedecker1,Goedecker2,Yang,Recursion,Krylov,DM,siesta,conquest,onetep,Varga,Tsuchida,Shimojo,Takayama,Ogura,Baer,Daniels,Kitaura}
have been devoted to extend applicability 
of the density functional theory (DFT)\cite{Hohenberg,Kohn} to large-scale systems,
which leads to realization of more realistic simulations being close to experimental 
conditions.
In fact, lots of large-scale DFT calculations have already contributed for 
comprehensive understanding of a vast range of materials,\cite{Gervasio,Miyazaki,Nishio,Zonias,Iwata,Choe}
although widely used functionals such as local density approximation (LDA)\cite{LDA} and 
generalized gradient approximation (GGA)\cite{GGA} have limitation in describing 
strong correlation in transition oxides and van der Waals interaction 
in biological systems. 

The efficient methods developed so far within the conventional 
DFT can be classified into two categories in terms of the computational complexity,
\cite{CP,Payne,Davidson,Pulay,Wood,Teter,Stich,Kresse,Householder,Golub,Goedecker1,Goedecker2,Yang,Recursion,Krylov,DM,siesta,conquest,onetep,Varga,Tsuchida,Shimojo,Takayama,Ogura}  
while the other methods, which deviate from the classification, have been also 
proposed.\cite{Baer,Daniels,Kitaura}
The first category consists of O($N^{3}$) methods,
\cite{CP,Payne,Davidson,Pulay,Wood,Teter,Stich,Kresse,Householder,Golub} 
where $N$ is the number of basis functions, as typified 
by the Householder-QR method,\cite{Householder,Golub} 
the conjugate gradient method,\cite{Payne,Teter,Stich} and 
the Pulay method,\cite{Pulay,Wood} 
which have currently become standard methods. 
The methods can be regarded as numerically exact methods, and 
the computational cost scales as O($N^{3}$) even if only valence states 
are calculated because of the orthonormalization process. 
On the other hand, the second category involves approximate O($N$) methods such as 
the density matrix method,\cite{DM,conquest,onetep} 
the orbital minimization method,\cite{siesta,Tsuchida}
and the Krylov subspace method\cite{Recursion,Krylov,Takayama} 
of which computational cost is proportional to the number of basis functions $N$. 
The linear-scaling of the computational effort in the O($N$) methods can be achieved
by introducing various approximations like the truncation of 
the density matrix\cite{DM} or Wannier functions\cite{siesta,Tsuchida} 
in real space. Although the O($N$) methods have been proven to be 
very efficient, the applications must be performed with careful consideration 
due to the introduction of the approximations, which might be one of reasons that 
the O($N$) methods have not been widely used compared to the O($N^{3}$) methods. 
From the above reason one may think of whether a numerically exact but low-order 
scaling method can be developed by utilizing the resultant sparse structure of 
the Hamiltonian and overlap matrices expressed by localized basis functions.
Recently, a direction towards the development of O($N^{2\sim}$) methods
has been suggested by Lin et al., in which diagonal elements of the density matrix 
is computed by a contour integration of the Green function calculated 
by making full use of the sparse structure of the matrix.\cite{Lin} 
Also, an efficient scheme has been presented by Li et al. to calculate 
diagonal elements of the Green function for electronic transport calculations,\cite{Li} 
which is based on the algorithm by Takahashi et al.\cite{Takahashi} 
and Erisman and Tinney.\cite{Erisman} 
However, except for the two methods mentioned above the development of 
numerically exact O($N^{2\sim}$) methods, which are positioned in between 
the O($N$) and O($N^{3}$) methods, has been 
rarely explored yet for large-scale DFT calculations.

In this paper we present a numerically exact but low-order scaling method 
for large-scale DFT calculations of insulators and metals using localized basis 
functions such as pseudo-atomic orbital (PAO),\cite{PAO} 
finite element (FE),\cite{FEM} and wavelet basis functions.\cite{Wavelet} 
The computational effort of the method scales as O($N(\log_2N)^2$), O($N^{2}$), 
and O($N^{7/3}$) for one, two, and three dimensional (1D, 2D, and 3D) systems, 
respectively. In spite of the low-order scaling, the method is a numerically 
exact alternative to the conventional O($N^{3}$) methods. 
The key idea of the method is to directly compute selected elements of the 
density matrix by a contour integration of the Green function evaluated with 
a set of recurrence formulas. 
It is shown that a contour integration method based on a continued fraction 
representation of the Fermi-Dirac function\cite{CF} can be successfully employed for
the purpose, and that the number of poles used in the contour 
integration does not depend on the size of the system.
We also derive a set of recurrence formulas based on the nested dissection\cite{George}
of the sparse matrix and a block $LDL^T$ factorization using the Schur complement\cite{Golub}
to calculate selected elements of the Green function. The computational complexity 
is governed by the calculation of the Green function. 
In addition to the low-order scaling, the method can be particularly advantageous 
to the massively parallel computation because of the well separated data structure.

This paper is organized as follows: 
In Sec.~II the theory of the proposed method is presented together with 
detailed analysis of the computational complexity.
In Sec.~III several numerical calculations are shown to illustrate practical 
aspects of the method within a model Hamiltonian and DFT calculations using 
the PAO basis functions. 
In Sec.~IV we summarize the theory and applicability of the numerically exact
but low-order scaling method.

\section{THEORY}
\subsection{Density matrix approach}

Let us assume that the Kohn-Sham (KS) orbital $\phi_{\nu}$ is expressed by a linear 
combination of localized basis functions $\{\chi\}$ such as PAO,\cite{PAO} FE,\cite{FEM} 
and wavelet basis functions\cite{Wavelet} as:
\begin{eqnarray}
  \phi_{\nu} ({\bf r}) = \sum_{i=1}^{N} c_{\nu i} \chi_{i}({\bf r}),\label{eqn:e1}
\end{eqnarray}
where $N$ is the number of basis functions. 
Throughout the paper, we consider the spin restricted and ${\bf k}$-independent 
KS orbitals for simplicity of notation. However, the generalization of our discussion
for these cases is straightforward.
By introducing LDA or GGA for the exchange-correlation functional, the KS equation 
is written in a sparse matrix form:
\begin{eqnarray}
  Hc_{\nu} = \varepsilon_{\nu} Sc_{\nu},\label{eqn:e2}
\end{eqnarray}
where $\varepsilon_{\nu}$ is the eigenvalue of state $\nu$, $c_{\nu}$ a vector 
consisting of coefficients $\{c_{\nu i}\}$, and $H$ and $S$ are 
the Hamiltonian and overlap matrices, respectively. 
Due to both the locality of basis functions and LDA or GGA for the
exchange-correlation functional, both the matrices possess the same 
sparse structure. It is also noted that the charge density $n({\bf r})$
can be calculated by the density matrix $\rho$:
\begin{eqnarray}
  n({\bf r}) = \sum_{i,j}\rho_{ij}\chi_{j}({\bf r})\chi_{i}({\bf r}).
  \label{eqn:e3}
\end{eqnarray}
By remembering that $\chi$ is localized in real space, one may notice 
that the product $\chi_i \chi_j$ is non-zero only if they are closely located 
each other. Thus, the number of elements in the density matrix required 
to calculate the charge density scales as O($N$).
As well as the calculation of the charge density, the total energy
is computed by only the corresponding elements of the density matrix
within the conventional DFT as: 
\begin{eqnarray}
  \nonumber
  E_{\rm tot}[n, \rho] &=& {\rm Tr}(\rho H_{\rm kin})
   + \int d{\bf r}n({\bf r }) v_{\rm ext}({\bf r })\\
   && + \int\int d{\bf r} d{\bf r}' 
  \frac{n({\bf r})n({\bf r}')}
       {\vert {\bf r}-{\bf r}'\vert} 
    + E_{\rm xc}[n],
\end{eqnarray}
where $H_{\rm kin}$ is the matrix for the kinetic operator, 
$v_{\rm ext}$ an external potential, and $E_{\rm xc}$ an exchange-correlation 
functional. Since the matrix $H_{\rm kin}$ possesses the same sparse structure 
as that of $S$, one may find an alternative way that
the selected elements of the density matrix, corresponding to the non-zero 
products $\chi_i \chi_j$, are directly computed without evaluating 
the KS orbitals. The alternative way enables us to avoid an orthogonalization
process such as Gram-Schmidt method for the KS orbitals, of which computational
effort scales as O($N^{3}$) even if only the occupied states are taken into account.
The direct evaluation of the selected elements in the density matrix  
is the starting point of the method proposed in the paper.
The density matrix $\rho$ can be calculated through the Green function $G$
as follows:
\begin{eqnarray}
  \rho = -\frac{2}{\pi}
   {\rm Im}\int_{-\infty}^{\infty}dE 
  G(E+i0^+) f\left(\frac{E-\mu}{k_{\rm B}T}\right),
\end{eqnarray}
where the factor 2 is due to the spin degeneracy, $f$ the Fermi-Dirac 
function, $\mu$ chemical potential, $T$ electronic temperature, $k_{\rm B}$
the Boltzmann factor, and $0^+$ a positive infinitesimal. 
Also the matrix expression of the Green function is given by 
\begin{eqnarray}
  G(Z) = (ZS-H)^{-1},
\end{eqnarray}
where $Z$ is a complex number.
Therefore, from Eqs.~(5) and (6), our problem is cast to two issues: 
(i) how the integration of the Green function can be efficiently performed,  
and 
(ii) how the selected elements of the Green function in the matrix form 
can be efficiently evaluated.
In the subsequent subsections we discuss the two issues in detail.

\subsection{Contour integration of the Green function}

We perform the integration of the Green function, Eq.~(5), by a contour 
integration method using a continued fraction representation of 
the Fermi-Dirac function.\cite{CF}
In the contour integration the Fermi-Dirac function is expressed by 
\begin{eqnarray}
  \nonumber
  \frac{1}{1+\exp(x)} 
  &=&
    \frac{1}{2} -  
    \frac{\strut \frac{x}{4}}
      {\displaystyle 1 + \frac{\strut (\frac{x}{2})^2}
      {\displaystyle 3 + \frac{\strut (\frac{x}{2})^2}
      {\displaystyle 5 + \frac{\strut (\frac{x}{2})^2}
      {\displaystyle \frac{\strut \cdots}
      {(2M-1)+}_{\ddots}
      }}}}\\
  &=&
   \frac{1}{2}
   +
   \sum_{p=1}^{\infty}\frac{R_p}{x-iz_p} 
   +
   \sum_{p=1}^{\infty}\frac{R_p}{x+iz_p},
\end{eqnarray}
where $x=\beta(Z-\mu)$ with $\beta=\frac{1}{k_{\rm B}T}$, 
$z_p$ and $R_p$ are poles of the continued fraction 
representation and the associated residues, respectively.  
The representation of the Fermi-Dirac function is derived from 
a hypergeometric function, and can be regarded as a Pad\'e approximant
when terminated at the finite continued fraction. 
The poles $z_p$ and residues $R_p$ can be easily obtained by
solving an eigenvalue problem as shown in Ref.~[\cite{CF}].
By making use of the expression of Eq.~(7) for Eq.~(5) and considering
the contour integration, one obtain the following expression for the integration
of Eq.~(5): 
\begin{eqnarray}
   \rho
    &=& 
    M^{(0)}
    + 
    {\rm Im}\left(
    -\frac{4i}{\beta} \sum_{p=1}^{\infty} G(\alpha_p)R_p
    \right),
\end{eqnarray}
where $\alpha_{p} = \mu + i\frac{z_p}{\beta}$, and $M^{(0)}$ is the zeroth 
order moment of the Green function which can be computed by 
$iRG(iR)$ with a large real number $R$.
The structure of the poles distribution,  that all the poles are located 
on the imaginary axis like the Matsubara pole, but the density of 
the poles becomes smaller as the poles go away from the real axis, 
has been found to be very effective for the efficient integration of 
Eq.~(5). It has been shown that only the use of the 100 poles at 600 K gives 
numerically exact results within double precision.\cite{CF} 
Thus, the contour integration method can be regarded as 
a numerically exact method even if the summation is terminated 
at a practically modest number of poles.

Moreover, it should be noted that the number of poles to achieve 
convergence is independent of the size of system. 
Giving the Green function in the Lehmann representation, Eq.~(8) can be 
rewritten by 
\begin{eqnarray}
   \nonumber
   \rho
    &=& 
    M^{(0)}
    + 
    {\rm Im}\left(
    -\frac{4i}{\beta} \sum_{p=1}^{\infty} 
    \sum_{\nu}
    \frac
     {\vert\phi_{\nu}\rangle\langle \phi_{\nu}\vert}
     {\alpha_p-\epsilon_{\nu}}
    R_p
    \right)\\
   &=&
    M^{(0)}
    + 
    \sum_{\nu}
    {\rm Im}\left(
    -\frac{4i}{\beta} \sum_{p=1}^{\infty} 
    \frac
     {\vert\phi_{\nu}\rangle\langle \phi_{\nu}\vert}
     {\alpha_p-\epsilon_{\nu}}
    R_p
    \right).
\end{eqnarray}
Although the expression in the second line is obtained by just exchanging 
the order of the two summations, the expression clearly shows that 
the number of poles for convergence does not depend on the size of system 
if the spectrum radius is independent of the size of system. 
Since the independence of the spectrum radius can be found in general cases, 
it can be concluded that the computational effort is determined by that for 
the calculation of the Green function.

The energy density matrix $e$, which is needed to calculate forces on atoms within 
non-orthogonal localized basis functions, can also be calculated by the 
contour integration method\cite{CF} as follows:
\begin{eqnarray}
  \nonumber
  e 
   &=&
  -\frac{2}{\pi}
   {\rm Im}\int_{-\infty}^{\infty}dE~E 
  G(E+i0^+) f\left(\frac{E-\mu}{k_{\rm B}T}\right),\\
  \nonumber
   &=&
  M^{(1)} 
  + 
  \kappa M^{(0)}
  + 
 {\rm Im}
 \left(
  -\frac{4i}{\beta}
  \sum_{p=1}^{\infty}
  G(\alpha_p)R_p \alpha_p 
 \right)\\ 
\end{eqnarray}
with $\kappa$ defined by
\begin{eqnarray}
  \kappa = \frac{4}{\beta}\sum_{p=1}^{\infty} R_p,
\end{eqnarray}
where $M^{(0)}$ and $M^{(1)}$ are the the zeroth and first
order moments of the Green function, and 
can be computed by solving 
the following simultaneous linear equation:
\begin{eqnarray}
  \left(
    \begin{array}{cc}
    1 & z_0^{-1}\\
    1 & z_1^{-1}
    \end{array}
  \right)
  \left(
    \begin{array}{c}
      M^{(0)}\\
      M^{(1)}
    \end{array}
  \right)
  = 
  \left(
    \begin{array}{c}
      z_0 G(Z_0)\\
      z_1 G(Z_1)\\
    \end{array}
  \right).
\end{eqnarray}
The equation is derived by terminating the summation over the order of 
the moments in the moment representation of the Green function.
By letting $z_0$ and $z_1$ be $iR$ and $-R$, respectively, $M^{(0)}$ 
and $M^{(1)}$ are explicitly given by 
\begin{eqnarray}
  M^{(0)} 
  &=&
 \frac{R}{1-i}
 \left(
   G(iR) - G(-R) 
 \right),\\
  M^{(1)} 
  &=&
 \frac{iR^2}{1+i}
 \left(
   iG(iR) + G(-R) 
 \right), 
\end{eqnarray}
where $R$ should be a large real number, and $10^7$ is used in this study 
so that the higher order terms 
can be negligible in terminating the summation in 
the moment representation of the Green function. 
Inserting Eqs.~(13) and (14) into Eq.~(10), we obtain the following expression 
which is suitable for the efficient implementation in terms of memory consumption:
\begin{eqnarray}
  \nonumber
  e 
   &=&
  \lambda G(iR) 
  + 
  \gamma G(-R)
  + 
 {\rm Im}
 \left(
  -\frac{4i}{\beta}
  \sum_{p=1}^{\infty}
  G(\alpha_p)R_p \alpha_p 
 \right)\\ 
\end{eqnarray}
with $\lambda$ and $\gamma$ defined by 
\begin{eqnarray}
  \lambda 
  &=&
  \frac{R}{2} 
  (1+i)(1+i\kappa R),\\
  \gamma
  &=&
  -\frac{R}{2} 
   (1+i)(1-\kappa R).
\end{eqnarray}
One may notice that the number of poles for convergence does not depend 
on the size of system even for the calculation of the energy density matrix 
because of the same reason as for the density matrix.

\subsection{Calculation of the Green function}

It is found from the above discussion that the computational effort to compute 
the density matrix is governed by that for the calculation of the Green function, 
consisting of an inversion of the sparse matrix of which computational effort by 
conventional schemes such as the Gauss elimination or LU factorization 
based methods scales as O($N^{3}$). Thus, the development of an efficient method
of inverting a sparse matrix is crucial for efficiency of the proposed method.

Here we present an efficient low-order scaling method, based on a nested dissection 
approach,\cite{George} of computing only selected elements in the inverse of a sparse 
matrix. The low-order scaling method proposed here consists of two steps: 
(1) {\it Nested dissection}: 
by noting that a matrix $X\equiv (ZS-H)$ is sparse, a structured matrix 
is constructed by a nested dissection approach. In practice, just reordering 
the column and row indices of the matrix $X$ yields the structured matrix. 
(2) {\it Inverse by recurrence formulas}:
by recursively applying a block $LDL^T$ factorization\cite{Golub} to the structured matrix, 
a set of recurrence formulas is derived. Using the recurrence formulas, 
only the selected elements of the inverse matrix $X^{-1}\equiv G(Z)$ are directly computed. 
The computational effort to calculate the selected elements in the inverse matrix
using the steps (i) and (ii) scales as O($N(\log_2N)^2$), O($N^{2}$), and O($N^{7/3}$) 
for 1D, 2D, and 3D systems, respectively, as shown later.
First, we discuss the nested dissection of a sparse matrix, and then derive 
a set of recurrence formulas of calculating the selected elements 
of the inverse matrix.

\begin{figure}[t]
    \centering
    \includegraphics[width=8.0cm]{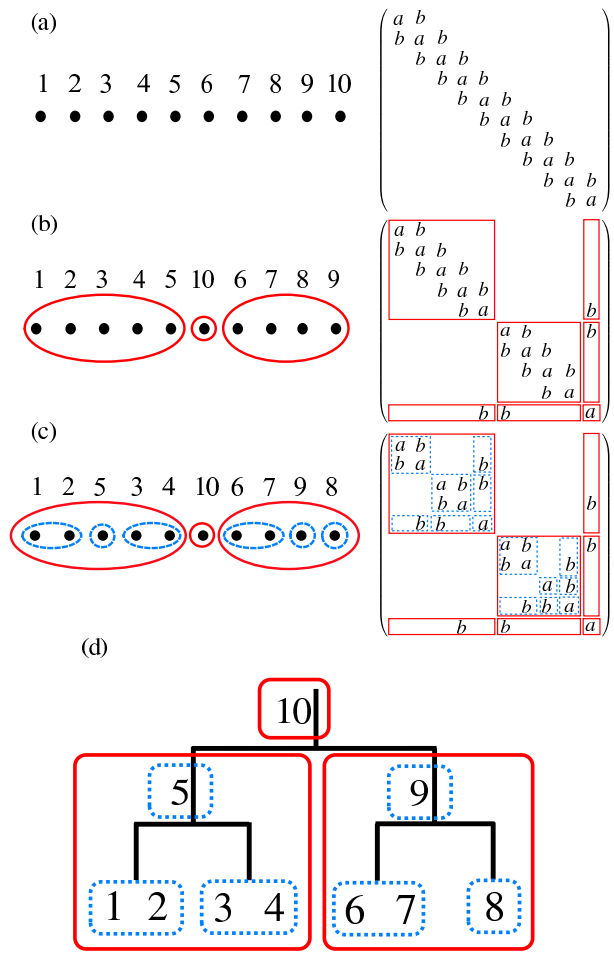}
    \caption{(Color online) 
     (a) The initial numbering for atoms in a linear chain molecule consisting 
         of ten atoms described by the $s$-valent NNTB and its corresponding matrix, 
     (b) the renumbering for atoms by the first step in the nested dissection and 
         its corresponding matrix, 
     (c) the renumbering for atoms by the second step in the nested dissection and 
         its corresponding matrix, 
     (d) the binary tree structure representing hierarchical interactions between 
         domains in the structured matrix by the numbering shown in Fig.~1(c). 
             }
\end{figure}

\subsubsection{Nested dissection}

As an example the right panel of Fig.~1(c) shows a structured matrix obtained 
by the nested dissection approach for a finite chain model consisting of ten atoms, 
where we consider a $s$-valent nearest neighbor tight binding (NNTB) model. 
When one assigns the number to the ten atoms as shown in the left panel 
of Fig.~1(a), then $X$ is a tridiagonal matrix, of which diagonal 
and off-diagonal terms are assumed to be $a$ and $b$, respectively, as shown 
in the right panel of Fig.~1(a). 
As the first step to generate the structured matrix
shown in the right panel of Fig.~1(c), we make a {\it dissection} of the system
into the left and right {\it domains}\cite{domain}
by renumbering for the ten atoms, and obtain
a dissected matrix shown in the right panel of Fig.~1(b). 
The left and right domains interact with each other through only  
a {\it separator} consisting of an atom 10. As the second step 
we apply a similar dissection for each domain generated by the first step, 
and arrive at a {\it nested}-{\it dissected} matrix given by the right 
panel of Fig.~1(c).  
The subdomains, which consist of atoms 1 and 2 and atoms 3 and 4, respectively, 
in the left domain
interact with each other through only a separator consisting of an atom 5. 
The similar structure is also found in the right domain consisting of atoms 
6, 7, 9, and 8. 
It is worth mentioning that the resultant nested structure of the sparse 
matrix can be mapped to a binary tree structure which indicates hierarchical 
interactions between (sub)domains as shown in Fig.~1(d). 
By applying the above procedure to a sparse matrix, one can convert any 
sparse matrix into a nested and dissected matrix in general. However in practice 
there is no obvious way to perform the nested dissection for general sparse
matrices, while a lot of efficient and effective methods have been already developed
for the purpose.\cite{Karypis,Davis}
Here we propose a rather simple but effective way for
the nested dissection by taking account of a fact that the basis function 
we are interested in is localized in real space, and that the sparse structure 
of the resultant matrix is very closely related to the position of basis 
functions in real space. The method bisects a system into two domains 
interacting through only a separator, and recursively applies to the 
resultant subdomains, leading to a binary tree structure for the interaction.
Our algorithm for the nested dissection of a general sparse matrix 
is summarized as follows:

(i) {\it Ordering}. 
Let us assume that there are $N_d$ basis functions in a domain 
we are interested in. We order the basis functions in the domain
by using the fractional coordinate for the central position of localized 
basis functions along ${\bf a}_i$-axis, 
where $i=1,2$, and 3. As a result of the ordering, each basis function can be
specified by the {\it ordering number}, which runs from 1 to $N_d$
in the domain of the central unit cell. The ordering number in the 
periodic cells specified by $l{\bf a}_i$, where $l=0,\pm 1, \pm 2,\cdots$, is given 
by $lN_d+q$, where $q$ is the corresponding ordering number in the 
central cell.
In isolated systems, one can use the Cartesian coordinate instead of the 
fractional coordinate without losing any generality.

(ii) {\it Screening of basis functions with a long tail}.
The basis functions with a long tail tend to make an efficient dissection 
difficult. The sparse structure formed by the other basis functions with 
a short tail is latescent due to the existence of the basis functions 
with a long tail. Thus, we classify the basis functions with 
a long tail in the domain as members in the separator before performing 
the dissection process. By the screening of the basis functions with 
a long tail, it is possible to expose concealed sparse structure 
when atomic basis functions with a variety of tails are used, 
while a systematic basis set such as the FE basis functions may 
not require the screening.

(iii) {\it Finding of a starting nucleus}. 
Among the localized basis functions in the domain, we search a basis 
function which has the smallest number of non-zero overlap with the 
other basis functions. Once we find the basis function, we set it 
as a starting {\it nucleus} of a subdomain. 

(iv) {\it Growth of the nucleus}.
Staring from a subdomain given by the procedure (iii), we grow 
the subdomain by increasing the size of nucleus step by step. 
The growth of the nucleus can be easily performed by managing 
the minimum and maximum ordering numbers, $m_{\rm min}$ and $m_{\rm max}$,
which ranges from 1 to $N_d$. We define the subdomain by basis 
functions with the successive ordering numbers between the 
minimum and maximum ordering numbers $m_{\rm min}$ and $m_{\rm max}$.
At each step in the growth of the subdomain, we search two basis 
functions which have the minimum ordering number $n_{\rm min}$ and maximum
ordering number $n_{\rm max}$ among basis functions overlapping with 
the subdomain defined at the growth step. 
In the periodic boundary condition, $n_{\rm min}$ can be smaller than zero, 
and $n_{\rm max}$ can be larger than the number of basis functions $N_d$. 
Then, the number of basis functions in the subdomain, the separator, 
and the other subdomain can be calculated by 
$N_0\equiv m_{\rm max}-m_{\rm min}+1$, 
$N_{\rm s}\equiv n_{\rm max}-n_{\rm min}+1-N_0$, 
and $N_{1}\equiv N_d-N_0-N_{\rm s}$, respectively, at each growth step.
By the growth process one can minimize $(\vert N_0-N_1\vert+N_{\rm s})$
being a measure for quality of the dissection, where 
the measure $(\vert N_0-N_1\vert+N_{\rm s})$ takes 
equal bisection size of the subdomains and minimization of the size 
of the separator into account.
Also, if $(n_{\rm max}-n_{\rm min}+1)$ is larger than $N_d$, then 
this situation implies that the proper dissection can be difficult
along the axis.  

(v) {\it Dissection}. 
By applying the above procedures (i)-(iv) to each ${\bf a}_i$-axis, 
where $i=1,2$, and 3, and we can find an axis which gives the minimum 
$(\vert N_0-N_1\vert+N_{\rm s})$. 
Then, the dissection along the axis is performed by renumbering for 
basis functions in the domain, and two subdomains and 
one separator are obtained. Evidently, the same procedures can be applied to 
each subdomain, and recursively continued until the size of 
domain reaches the threshold. 
As a result of the recursive dissection, we obtain a structured 
matrix by the nested dissection. 

As an illustration we apply the method for the nested dissection
to the finite chain molecule shown in Fig.~1. 
We first set all the system as {\it domain}, and start to apply 
the series of procedures to the domain.  
The procedure (i) is trivial for the case, and we obtain the numbering 
of atoms and the corresponding matrix shown in Fig.~1(a). 
Also it is noted that the screening of 
the basis functions with a long tail is unnecessary, and that we only have 
to search the chain direction.
In the procedure (iii), atoms 1 and 10 in Fig.~1(a) satisfy the condition.
Choosing the atom 1 as a starting nucleus of the domain, and we 
gradually increase the size of the domain according to the 
procedure (iv). Then, it is found that the division shown in Fig.~1(b)
gives the minimum $(\vert N_0-N_1\vert+N_{\rm s})$. 
Renumbering for the basis functions based on the analysis yields 
the dissected matrix shown in the right panel of Fig.~1(b). 
By applying the similar procedures to the left and right subdomains, 
one will immediately find the result of Fig.~1(c). 
\begin{figure}[t]
    \centering
    \includegraphics[width=8.8cm]{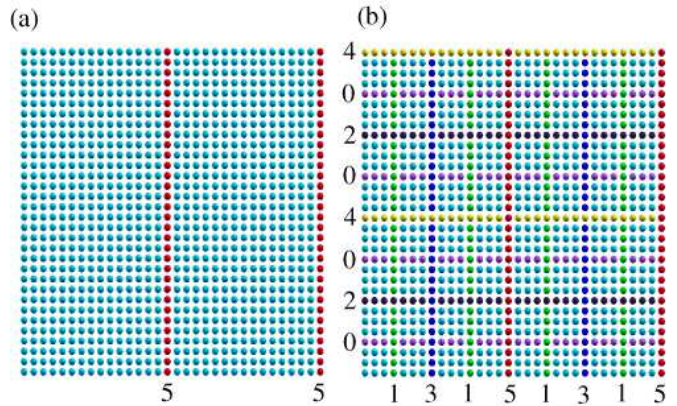}
    \caption{(Color online) 
        (a) 
        The square lattice model, described by the $s$-valent NNTB, of which  
        unit cell contains 1024 atoms with periodic boundary condition. 
        The right blue and red circles correspond to atoms in two domains and 
        a separator, respectively, at the first step in the nested dissection.
        (b) The square lattice model at the final step in the nested dissection. 
        The separator at the innermost and the outermost levels are labeled as
        separators 0 and 5, respectively, and the separators at each level are  
        constructed by atoms with a same color.
         }
\end{figure}
In addition to the finite chain molecule, as an example of more general cases, 
the above algorithm for the nested dissection is applied to a $s$-valent NNTB 
square lattice model of 
which unit cell contains 1024 atoms with periodic boundary condition.
At the first step in the nested dissection, the separator is found 
to be red atoms as shown in Fig.~2(a). Due to the periodic boundary condition, 
the separator consists of two {\it lines}. At the final step, the system 
is dissected by the recursive algorithm as shown in Fig.2~(b). 
The separator at the innermost and the outermost levels are labeled as
separators 0 and 5, respectively, and each subdomain at the innermost level
includes 9 atoms. As demonstrated for the square lattice model, 
the algorithm can be applied for systems with any dimensionality, and provides 
a well structured matrix for our purpose in a single framework.

\subsubsection{Inverse by recurrence formulas}

We directly compute the selected elements of the inverse matrix using 
a set of recurrence formulas which can be derived by recursively applying 
a block $LDL^T$ factorization to the structured matrix obtained by the nested 
dissection method as shown below.  
To derive the recurrence formulas, we first introduce the block $LDL^T$ 
factorization\cite{Golub} for a symmetric square matrix $X$:  
\begin{eqnarray}
  \nonumber
  X &=& 
  \left(
    \begin{array}{cc}
    A & B^{T}\\
    B & C
    \end{array}
  \right)\\
  &=& 
  \left(
    \begin{array}{cc}
    I & 0\\
    L & I
    \end{array}
  \right)
  \left(
    \begin{array}{cc}
    A & 0\\
    0 & S
    \end{array}
  \right)
  \left(
    \begin{array}{cc}
    I & L^{T}\\
    0 & I
    \end{array}
  \right),
\end{eqnarray}
where $A$ and $C$ are diagonal block matrices, and $B$ and $B^{T}$ 
an off-diagonal block matrix and its transposition, and $L$ is given by 
\begin{eqnarray}
  L = B A^{-1}. 
\end{eqnarray}
Also the Schur complement $S$ of the block element $C$ is defined by 
\begin{eqnarray}
  S \equiv C - B A^{-1} B^{T} = C - BL^{T}. 
\end{eqnarray}
Then, it is verified that the inverse matrix of $X$ is given by 
\begin{eqnarray}
  X^{-1} = 
  \left(
    \begin{array}{cc}
    A^{-1}+L^{T}S^{-1}L & -L^{T}S^{-1}\\
    -S^{-1}L & S^{-1}
    \end{array}
  \right).
\end{eqnarray}
We now consider calculating the selected elements of the inverse of the 
structured matrix given in Fig.~1(c) using Eq.~(21), and rewrite 
the matrix in Fig.~1(c) in a block form as follows:
\begin{eqnarray}
  X =  
  \left(
    \begin{array}{ccccccc}
    A_{0,0} &  & B_{0,0}^{T} & & & & \\
     & A_{0,1} & B_{0,1}^{T} & & & & B_{1,0}^{T} \\
    B_{0,0} & B_{0,1} & C_{0,0} & & & & \\
    & & & A_{0,2} &  & B_{0,2}^{T} & \\
    & & &  & A_{0,3} & B_{0,3}^{T} &  B_{1,1}^{T} \\
    & & & B_{0,2} & B_{0,3} & C_{0,1} & \\
    & B_{1,0} & & & B_{1,1} &  & C_{1,0} \\
    \end{array}
  \right),
  \label{eqn:e14}
\end{eqnarray}
where $A_{0,0}$ and $B_{0,0}$ correspond to 
$
  \left(
    \begin{array}{cc}
      a & b\\
      b & a\\
    \end{array}
  \right)
$ 
and 
$(0,b)$, respectively, and
the other block elements can be deduced. Also the blank indicates a block zero 
element.
Using Eq.~(20) the Schur complement of $C_{1,0}$ is given by 
\begin{eqnarray}
  S_{1,0} = C_{1,0} 
   -B_{1,0}
    L_{1,0}^{T}
   -B_{1,1}
    L_{1,1}^{T},
\end{eqnarray}
where $L^{T}_{1,0}$ is calculated by Eq.~(19) and can be 
transformed using Eq.~(21) to a recurrence formula as follows:
\begin{eqnarray}
  \nonumber
  L^{T}_{1,0} 
  &=& 
  \left(
    \begin{array}{ccccccc}
    A_{0,0} &  & B_{0,0}^{T} \\
     & A_{0,1} & B_{0,1}^{T} \\
    B_{0,0} & B_{0,1} & C_{0,0} \\
    \end{array}
  \right)^{-1}
   B_{1,0}^{T}\\
  \nonumber
  &=&
  \left(
    \begin{array}{ccccccc}
    A_{0,0}^{-1}
    & 
    & 
    \\
    & 
    A_{0,1}^{-1}
    & 
    \\
    &  
    & 0 \\
    \end{array}
  \right)
   B_{1,0}^{T}\\
  \nonumber
  &+&
  \left(
    \begin{array}{ccccccc}
    L^{T}_{0,0}S^{-1}_{0,0}L_{0,0} 
    & 
    L^{T}_{0,0}S^{-1}_{0,0}L_{0,1} 
    & 
   -L^{T}_{0,0} S_{0,0}^{-1} \\
    L_{0,1}^{T}S^{-1}_{0,0}L_{0,0}
    & 
    L^{T}_{0,1}S^{-1}_{0,0}L_{0,1} 
    & 
   -L^{T}_{0,1} S_{0,0}^{-1}\\
   -S_{0,0}^{-1}L_{0,0} 
    &  
   -S_{0,0}^{-1}L_{0,1} 
   & S_{0,0}^{-1} \\
    \end{array}
  \right)
   B_{1,0}^{T}\\
  &=&
  \left(
    \begin{array}{c}
     V_{1,0,0}^{T}\\
     V_{1,0,1}^{T}\\
     0\\
    \end{array}
  \right)
  + 
  \left(
    \begin{array}{c}
     L_{0,0}^{T}\\
     L_{0,1}^{T}\\
     -I\\
    \end{array}
  \right)Q_{1,1,0}^{T}
  \equiv 
  V_{1,1,0}^{T}
\end{eqnarray}
with the definitions: 
\begin{eqnarray}
  V_{1,0,0}^{T} &=& A_{0,0}^{-1}(B_{1,0}[B_{0,0}])^{T},\\
  V_{1,0,1}^{T} &=& A_{0,1}^{-1}(B_{1,0}[B_{0,1}])^{T},
\end{eqnarray}
and 
\begin{eqnarray}
  \nonumber
  Q_{1,1,0}^{T} = 
  S_{0,0}^{-1} 
  \left(
  B_{0,0}V_{1,0,0}^{T}
 +B_{0,1}V_{1,0,1}^{T} 
 -(B_{1,0}[C_{0,0}])^{T}  
  \right).\\
\end{eqnarray}
In Eqs.~(25), (26), and (27), we used a bra-ket notation [~] which 
stands for a part of the block element. For example, $B_{1,0}[B_{0,0}]$ 
means a part of $B_{1,0}$ which has the same columns as those of $B_{0,0}$.
It is noted that one can obtain a similar expression for $L^{T}_{1,1}$
as well as Eq.~(24) for $L^{T}_{1,0}$. 

To address a more general case where the dissection for the sparse matrix 
is further nested, we suppose that the matrix $A_{0,0}$ has the same inner 
structure as  
\begin{eqnarray}
  \nonumber
  \left(
    \begin{array}{ccccccc}
    A_{0,0} &  & B_{0,0}^{T} \\
     & A_{0,1} & B_{0,1}^{T} \\
    B_{0,0} & B_{0,1} & C_{0,0} \\
    \end{array}
  \right),
\end{eqnarray}
then one may notice the recursive structure in Eq.~(24), and 
can derive the following set of recurrence relations for general cases:
\begin{eqnarray}
 \nonumber
 && Q_{p,m+1,n}^{T} =
   S_{m,n}^{-1}\times\\
 \nonumber
   &&
   \left(
    B_{m,2n}V_{p,m,2n}^{T}
   +B_{m,2n+1}V_{p,m,2n+1}^{T}
   -(B_{p,q}[C_{m,n}])^T
   \right),  \label{eqn:e20}\\\\
  \nonumber
 &&
  V_{p,m+1,n}^{T} =
  \left(
    \begin{array}{c}
     V_{p,m,2n}^{T} 
     \\
     V_{p,m,2n+1}^{T} 
     \\
     0 
    \end{array}
  \right)
  +
  \left(
    \begin{array}{c}
    L_{m,2n}^{T}
     \\
    L_{m,2n+1}^{T}
     \\
   -I
    \end{array}
  \right) Q_{p,m+1,n}^{T}.\\
  \label{eqn:e21}
\end{eqnarray}
Equation (\ref{eqn:e21}) is the central recurrence formula coupled with Eq.~(\ref{eqn:e20}), 
where the initial block elements are given by 
\begin{eqnarray}
  V_{p,0,n}^{T} = (A_{0,n})^{-1} (B_{p,q}[B_{0,n}])^{T}.
  \label{eqn:e22}
\end{eqnarray}
Also $L_{p,n}$ and $S_{p,n}$ can be calculated by 
\begin{eqnarray}
 && L_{p,n} = V_{p,p,n},\label{eqn:e23}\\
 && S_{p,n} = C_{p,n} 
 -(B_{p,2n},B_{p,2n+1})
  \left(
    \begin{array}{c}
     L_{p,2n}^{T} 
     \\
     L_{p,2n+1}^{T}
     \\ 
    \end{array}
  \right).\label{eqn:e24}
\end{eqnarray}
A set of Eqs.~(\ref{eqn:e20})-(32) enables us to calculate all the inverses of 
the Schur complements $S$ and $L$.
In the recurrence equations Eqs.~(\ref{eqn:e20}) and (\ref{eqn:e21}), three 
indices of $p$, $m$, and $n$ are involved, and they run as follows: 
\begin{eqnarray}
  p &=& 0, \cdots, P.\\
  m &=& 0, \cdots, p-1.\\
  n &=& 0, \cdots, 2^{P-m}-1.
\end{eqnarray}
The index $p$ denotes the level of hierarchy in the nested dissection and 
the innermost and outermost levels are set to 0 and $P$, respectively.
Then, it is noted that the total system is divided into $2^{P+1}$ domains 
at the innermost level. As well as $p$ the index $m$ is also related to 
the level of hierarchy in the nested dissection, and runs from 0 to $p-1$.
The index $n$ is a rather intermediate one, being dependent on $m$.
The indices $n$ in Eq.~(30) is dependent on $p$ and $q$ and they run 
as follows:
\begin{eqnarray}
  n&=&q(2^p),\cdots,(q+1)(2^p)-1.\\
  q&=&0,\cdots, 2^{P+1-p}-1.
\end{eqnarray}
Since the set of the recurrence formulas Eqs.~(\ref{eqn:e20})-(32) proceed according to 
Eqs~.(33)-(35), the development of recurrence can be illustrated as in Fig.~3. 
The recurrence starts from Eq.~(30) with $p=0$, and Eqs.~(31) and (32) follow. 
Then, $p$ is incremented by one, and $m+1$ climbs up to 1. 
The increment of $p$ and the climbing 
of $m+1$ are repeated until $p=P$ and $m+1=P$.
At $m+1=p$ for each $p$, $L$ and $S$ are evaluated by Eqs.~(31) and (32), 
and the inverse of $S$ is calculated by a conventional method such as LU factorization, 
which are used in the next recurrence for the higher level of hierarchy.
The numbers in the right hand side of Fig.~3 give the multiplicity for similar 
calculations by Eq.~(\ref{eqn:e21}) coming 
from the index $n$ at each $m+1$, since $n$ runs from 0 to $2^{P-m}-1$ as given 
in Eq.~(35). The computational complexity can be estimated by Fig.~3, and we will 
discuss its details later. 

\begin{figure}[t]
    \centering
    \includegraphics[width=8.5cm]{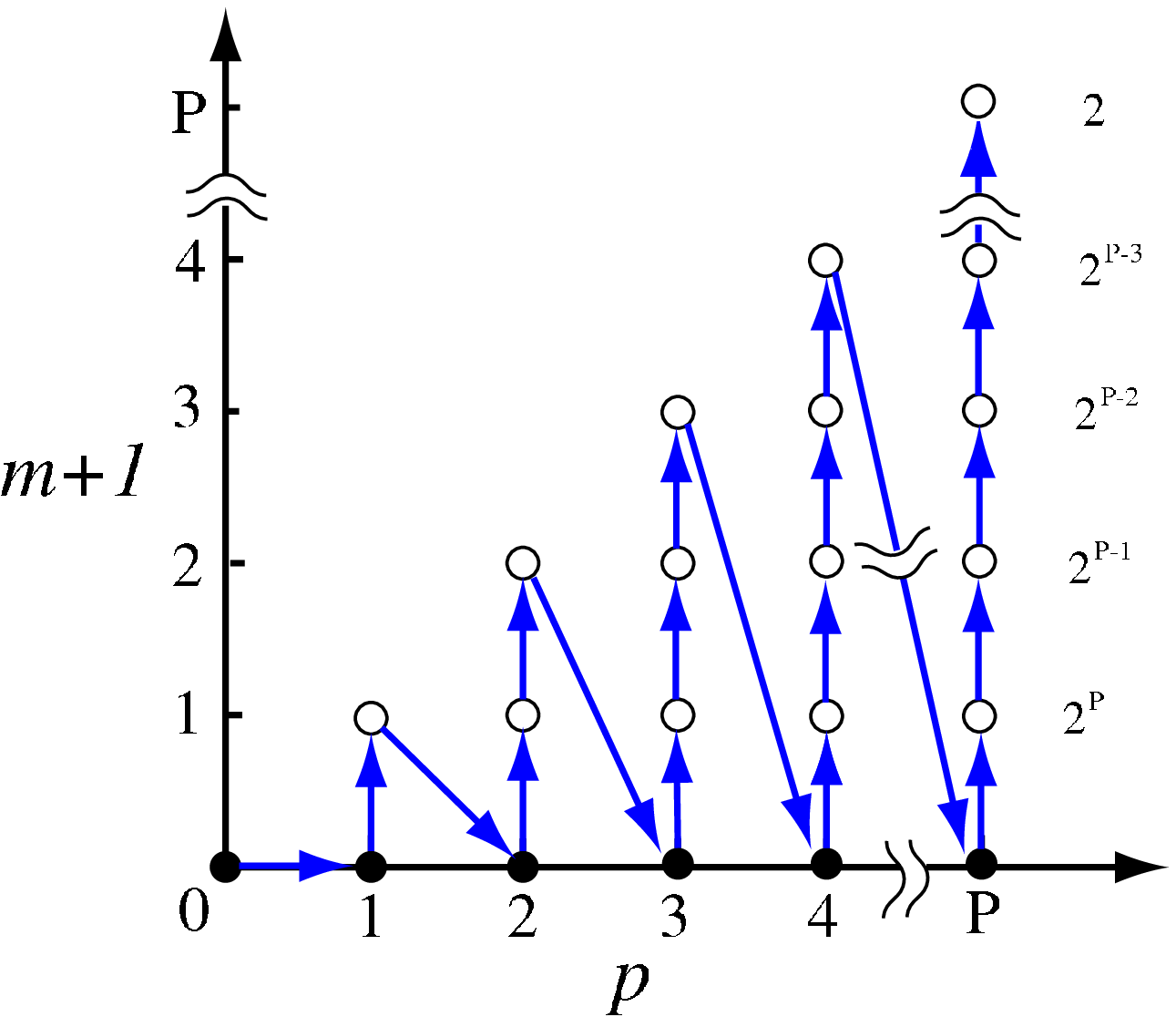}
    \caption{(Color online) 
            The development of recurrence formulas Eqs.~(\ref{eqn:e20})-(32), which 
            implies that the recurrence starts from $p=m+1=0$ and ends at 
            $p=m+1=P$. The number in the right hand side is the multiplicity 
            for similar calculations by Eq.~(\ref{eqn:e21}) due to the index $n$ 
            at each $m+1$. 
            }
\end{figure}

We are now ready to calculate the selected elements of the Green function using 
the inverses of the Schur complements $S$ and $L$ calculated by the recurrence 
formulas of Eqs.~(\ref{eqn:e20})-(32). By noting that Eq.~(21) has a recursive structure and 
the matrix $X$ is structured by the nested dissection, one can derive the following 
recurrence formula:
\begin{eqnarray}
 \nonumber
  && X^{-1}_{p+1,n} = 
  \left(
    \begin{array}{ccc}
    X_{p,2n}^{-1} & & \\
     & X_{p,2n+1}^{-1} & \\
     & & 0
    \end{array}
  \right)\\
  && 
  + 
  \left(
    \begin{array}{ccc}
    Y_{p,2n}^{T}L_{p,2n} & & -Y_{p,2n}^{T}\\
     & Y_{p,2n+1}^{T}L_{p,2n+1} & -Y_{p,2n+1}^{T}\\
    -Y_{p,2n} & -Y_{p,2n+1} & S^{-1}_{p,n}
    \end{array}
  \right),\label{eqn:e30}
\end{eqnarray}
where 
\begin{eqnarray}
  \nonumber
  Y_{p,2n}^{T} &=& L^{T}_{p,2n}S^{-1}_{p,n},\\ 
  Y_{p,2n+1}^{T} &=& L^{T}_{p,2n+1}S^{-1}_{p,n}.\label{eqn:e31} 
\end{eqnarray}
The recurrence formula Eq.~(38) starts with $X_{0,n}^{-1}=(A_{0,n})^{-1}$, 
adds contributions at $m+1=p$ for every $p$, and at last yields the inverse of the matrix 
$X$ as $X^{-1}=G(Z)=X_{P+1,0}^{-1}$. Since the calculation of each element for the inverse 
of $X$ can be independently performed, only the selected elements can be computed without 
calculating all the elements. The selected elements to be calculated are elements in 
the block matrices $A$, $B$, and $C$, each of which corresponds to a non-zero overlap 
matrix as discussed before. Thus, we can easily compute only the selected elements
using a table function which stores the position for the non-zero elements in the block 
matrices $A$, $B$, and $C$.

A simple but nontrivial example is given in Appendix A to illustrate 
how the inverse of matrix is computed by the recurrence formulas, and also 
a similar way is presented to calculate a few eigenstates around a selected 
energy in Appendix B, while the proposed method can calculate the total energy
of system without calculating the eigenstates.

\subsubsection{Finding chemical potential}

As well as the conventional DFT calculations, in the proposed method the chemical potential 
has to be adjusted so that the number of electrons can be conserved. However, there is no 
simpler way to know the number of electrons under a certain chemical potential before the 
contour integration by Eq.~(8) with the chemical potential. Thus, we search the chemical 
potential by iterative methods for the charge conservation. Since the contour integration
is the time-consuming step in the method, a smaller number of the iterative step directly 
leads to the faster calculation. Therefore, we develop a careful combination of several 
iterative methods to minimize the number of the iterative step for sufficient convergence.
In general, the procedure for searching the chemical potential can be performed by a sequence 
(1)-(2) or (5)-(1)-(3)-(1)-(4)-(1)-(4)-(1)$\cdots$ in terms of the following procedures.
As shown later, the procedure enables us to obtain the chemical potential conserving 
the number of electrons within $10^{-8}$ electron/system by less than 5 iterations on an average.

(1) {\it Calculation of the difference $\Delta N_0$ in the total number of electrons}.
The difference $\Delta N_i$ in the total number of electrons is defined 
with $\rho(\mu_i)$ calculated using Eq.~(8) at a chemical potential 
$\mu_i$ by
\begin{eqnarray}
  \Delta N_i = {\rm Tr}\left(\rho(\mu_i) S\right) - N_{\rm ideal},
\end{eqnarray}
where $N_{\rm ideal}$ is the number of electrons that the system should possess for 
the charge conservation. If $\Delta N_0$ is zero, the chemical potential 
$\mu_0$ is the desired one of the system.

(2) {\it Using the retarded Green function}.
If the difference $\Delta N_0$ is large enough so that the interpolation schemes 
(3) and (4) can fail to guess a good chemical potential, 
the next trial chemical potential is estimated by using the retarded Green function.
When the chemical potential of $\mu_{\rm tri}$ is considered, 
the correction $\delta N_{\rm tri}$ to $\Delta N_i$ estimated by the retarded 
Green function is given by 
\begin{eqnarray}
  \delta N_{\rm tri} =
  \int_{E_{\rm min}}^{E_{\rm max}} 
   dE \delta \rho(E) \Delta f(E,\mu_{\rm tri}),    
\end{eqnarray}
where $\delta \rho(E)$ and $\Delta f(E,\mu_{\rm tri})$ are defined by 
\begin{eqnarray}
  \delta \rho(E) = -\frac{2}{\pi}
      {\rm Im} 
      {\rm Tr}\left(
           G(E + i\eta) S
           \right)
\end{eqnarray}
with a small number $\eta$ (0.01~eV in this study) and 
\begin{eqnarray}
   \Delta f(E,\mu_{\rm tri} ) 
   &=& 
   f\left(\frac{E-\mu_{\rm tri}}{k_{\rm B}T}\right)
   -
   f\left(\frac{E-\mu_i}{k_{\rm B}T}\right).
\end{eqnarray}
The integration in Eq.~(41) is numerically evaluated by a simple quadrature 
scheme such as trapezoidal rule with a similar number of points as for that of 
poles in Eq.~(8), and the integration range can be determined
by considering the surviving range of $\Delta f(E,\mu_{\rm tri})$.  
The search of $\mu_{\rm tri}$ is performed by a bisection method
until $\Delta N_{\rm cri}>(\Delta N_i+\delta N_{\rm tri})$, where 
$\Delta N_{\rm cri}$ is a criterion for the convergence, and 
$10^{-8}$ electron/system is used in this study. It should be noted that 
the evaluation of Green function being the time-consuming part can be 
performed before the bisection method and a set of $\delta \rho(E)$ is stored 
for computational efficiency.

(3) {\it Linear interpolation/extrapolation method}.
In searching the chemical potential $\mu$, if two previous results ($\mu_i, \Delta N_i$) and ($\mu_j, \Delta N_j$)
are available, a trial chemical potential $\mu_{\rm tri}$ is estimated
by a linear interpolation/extrapolation method as: 
\begin{eqnarray}
  \mu_{\rm tri} = \frac{\mu_j \Delta N_i - \mu_i \Delta N_j}{\mu_i - \mu_j}.
\end{eqnarray}

(4) {\it Muller method}\cite{Muller,Wu}.
In searching the chemical potential $\mu$, if tree 
previous results ($\mu_i, \Delta N_i$), ($\mu_j, \Delta N_j$), 
and ($\mu_k, \Delta N_k$) are available, they can be fitted to a 
quadratic equation:
\begin{eqnarray}
  \Delta N = a\mu^2 + b\mu + c,
\end{eqnarray}
where $a$, $b$, and $c$ are found by solving a simultaneous linear equation
of $3\times 3$ in size.\cite{Muller_add}  
Then, $\mu_{\rm tri}$ giving $\Delta N=0$ is a solution of Eq.~(45), and given by 
\begin{eqnarray}
  \mu_{\rm tri} = 
  \left\{
  \begin{array}{ll}
   \frac{-2c}{b+\sqrt{b^2-4ac}} & b \geq 0,\\
   \frac{-b+\sqrt{b^2-4ac}}{2a} & b< 0.   
  \end{array}
  \right.
\end{eqnarray}
The selection of sign is unique because of the condition that the gradient
at the solution must be positive, and the branching is taken into account
to avoid the round-off error. 
As the iteration proceeds in search of the chemical potential, 
we have a situation that the number of available previous results 
is more than three. For the case, it is important to select three chemical 
potentials having smaller $\Delta N$ {\it and} the different sign of $\Delta N$ 
among three chemical potentials, since the guess of $\mu_{\rm tri}$ can be 
performed as the interpolation rather than the extrapolation. 

(5) {\it Extrapolation of chemical potential for the second step}.
During the self-consistent field (SCF) iteration, the chemical 
potential obtained at the last SCF step is used as the initial guess 
$\mu_1$ in the current SCF step. In addition, we estimate the second trial 
chemical potential by fitting results 
$(\mu_1^{(i)}, \Delta N_1^{(i)}, \mu_2^{(i)}, \Delta N_2^{(i)})$,
$(\mu_1^{(j)}, \Delta N_1^{(j)}, \mu_2^{(j)}, \Delta N_2^{(j)})$,
and 
$(\mu_1^{(k)}, \Delta N_1^{(k)}, \mu_2^{(k)}, \Delta N_2^{(k)})$, 
where the subscript and the superscript in $\mu_0^{(i)}$ and $\Delta N_0^{(i)}$ mean
the iteration step in search of the chemical potential and the SCF step, 
respectively,
at three previous SCF steps to the following equation:
\begin{eqnarray}
  \Delta N_2 = a_1 \Delta N_1 + a_2 (\mu_2-\mu_1) + a_3 \Delta N_1 (\mu_2-\mu_1),
\end{eqnarray}
where $a_1$, $a_2$, and $a_3$ are found by solving a simultaneous linear equation
of $3\times 3$ in size. Then, the chemical potential $\mu_2$ giving $\Delta N_2=0$
can be estimated by solving Eq.~(47) with respect to $\mu_2$ as follows:
\begin{eqnarray}
   \mu_{\rm tri} \equiv \mu_2 = \mu_1 - \frac{a_1 \Delta N_1}{a_2+a_3\Delta N_1}.
\end{eqnarray}
It is found from numerical calculations that Eq.~(48) provides a very accurate 
guess in most cases as the SCF calculation converges.

   \begin{table}[t]
     \caption{
        Some of $N^{(\rm 2)}_{m}$ and $N^{(\rm 3)}_{p}$ in Eq.~(50) 
        for a finite 1D chain, a finite 2D square lattice, 
        and a finite 3D cubic lattice described by the $s$-valent NNTB model. 
        They depends on $m$ or $p$ for the 2D and 3D systems in a rather complicated way,
        while $N^{(\rm 1)}_{p,m,n}=\frac{N}{2^{P-m}}$ for all the cases.
        The unit for each case is given in parenthesis.
       }
   \vspace{1mm}
   \begin{tabular}{lccccccccccc}
   \hline\hline
     m+1 or p & P & P-1 & P-2 & P-3 & P-4 & P-5 & P-6 & P-7 & P-8 & P-9 & P-10 \\
     \hline
      1D (1)        & 1 & 1   & 1   & 1 & 1 & 1 & 1 & 1 & 1 & 1 & 1\\
      2D ($N^{1/2}$)& 1 & $\frac{1}{2}$ & $\frac{1}{2}$ 
         & $\frac{1}{4}$ &  $\frac{1}{4}$ 
         & $\frac{1}{8}$ &  $\frac{1}{8}$
         & $\frac{1}{16}$ &  $\frac{1}{16}$
         & $\frac{1}{32}$ & $\frac{1}{32}$ \\ 
      3D ($N^{2/3}$)& 1 & $\frac{1}{2}$ & $\frac{1}{4}$ 
         & $\frac{1}{4}$ & $\frac{1}{8}$ 
         & $\frac{1}{16}$ & $\frac{1}{16}$ 
         & $\frac{1}{32}$ & $\frac{1}{64}$ 
         & $\frac{1}{64}$ & $\frac{1}{128}$ \\
   \hline
   \end{tabular}
  \end{table}

\subsection{Computational complexity}

We analyze the computational complexity of the proposed method.
As discussed in the subsection {\it Contour integration of the Green function},
the number of poles for the contour integration is independent of the size of system.
Thus, we focus on the computational complexity of the calculation of the Green function. 
For simplicity of the analysis we consider a finite chain, a finite square lattice, 
and a finite cubic lattice as representatives of 1D, 2D, and 3D systems, respectively, 
which are described by the $s$-valent NNTB models
as in the explanation of the nested dissection. Note that the results in the analysis 
are valid for more general cases with periodic boundary conditions. 
Since the computational cost is governed by Eq.~(\ref{eqn:e21}), let us first analyze the 
computational cost of Eq.~(\ref{eqn:e21}), while those of the other equations will be discussed later.
Considering that the recurrence formula of Eq.~(\ref{eqn:e21}) 
develops as shown in Fig.~3, and that the calculation of Eq.~(\ref{eqn:e21}) corresponds to 
the open circle in the figure, the computational cost $t$ can be estimated by 
\begin{eqnarray}
  t \propto \sum_{p=1}^{P} \sum_{m=0}^{p-1}\sum_{n=0}^{2^{P-m}-1} 
  N^{(\rm 1)}_{m} N^{(\rm 2)}_{m}N^{(\rm 3)}_{p},
\end{eqnarray}
where $N^{(1)}_{m}$ and $N^{(2)}_{m}$  are the dimension of 
row and column in the matrix: 
\begin{eqnarray}
  \nonumber
  \left(
    \begin{array}{c}
    L_{m,2n}^{T}
     \\
    L_{m,2n+1}^{T}
     \\
   -I
    \end{array}
  \right),
\end{eqnarray}
and $N^{(3)}_{p}$ is the dimension of column in the matrix $Q_{p,m+1,n}^{T}$.
Since Eq.~(\ref{eqn:e21}) consists of a matrix product, the computational cost is simply given 
by $N^{(\rm 1)}_{m} N^{(\rm 2)}_{m}N^{(\rm 3)}_{p}$. 
Also it is noted that $N^{(\rm 1)}_{m}$ and $N^{(\rm 2)}_{m}$ depend on 
only $m$, and $N^{(\rm 3)}_{p}$ has dependency on only $p$ because of 
the simplicity of the systems we consider.
 
For the finite 1D chain system, we see that 
$N^{(\rm 1)}_{m}=N/(2^{P-m})$ and $N^{(\rm 2)}_{m}=N^{(\rm 2)}_{p}=1$ as listed in 
Table I. 
Thus, the computational cost $t_{\rm 1D}$ for the 1D system is estimated as
\begin{eqnarray}
  \nonumber
  t_{\rm 1D} 
  &\propto& 
  \sum_{p=1}^{P} \sum_{m=0}^{p-1}\sum_{n=0}^{2^{P-m}-1} 
  \frac{N}{2^{P-m}}\\
  &=&
  \frac{1}{2}N P(P+1).
\end{eqnarray}
Noting $N\propto 2^P$, we see that the computational cost for the 1D system scales 
as O$(N (\log_2(N))^2)$.

For the finite 2D square lattice system, we see 
$N^{(\rm 1)}_{m}=N/(2^{P-m})$, and 
$N^{(\rm 2)}_{m}$ and $N^{(\rm 3)}_{p}$ depend on $m$ and $p$, respectively 
as shown in Table I. To estimate the order of the computational cost 
we approximate $N^{(\rm 2)}_{m}$ and $N^{(\rm 3)}_{p}$ as 
$N^{(\rm 2)}_{m}\approx N^{1/2}/2^{\frac{1}{2}(P-m-1)}$ and 
$N^{(\rm 3)}_{p}\approx N^{1/2}/2^{\frac{1}{2}(P-p)}$ which are equal to or more 
than the corresponding exact number.
Then, the computational cost $t_{\rm 2D}$ for the 2D system can be 
estimated as follows: 
\begin{eqnarray}
  \nonumber
  t_{\rm 2D} 
  &\propto& 
  \sum_{p=1}^{P} \sum_{m=0}^{p-1}\sum_{n=0}^{2^{P-m}-1} 
  \frac{N}{2^{P-m}}
   N^{(\rm 2)}_{p,m,n} 
   N^{(\rm 3)}_{p,m,n}\\
  \nonumber
  &<&
  \sum_{p=1}^{P} 
  \sum_{m=0}^{p-1}
  \sum_{n=0}^{2^{P-m}-1}
  \frac{N}{2^{P-m}}
  \frac{N^{1/2}}{2^{\frac{1}{2}(P-m-1)}}
  \frac{N^{1/2}}{2^{\frac{1}{2}(P-p)}}\\
 \nonumber
  &=&
  \frac{2N^2}{(\sqrt{2}-1)^2}
  \left(
    2-\sqrt{2} + \frac{\sqrt{2}}{2^P}-\frac{1}{2^P}-\frac{1}{2^{P/2}}
  \right).\\
\end{eqnarray}
Since the first twos term in parenthesis of the last line are the leading term, 
we see that the computational cost for the 2D system scales as O$(N^2)$.

   \begin{table}[t]
     \caption{
       Computational order of Eqs.~(\ref{eqn:e20}), (\ref{eqn:e21}), (\ref{eqn:e22}),
       (\ref{eqn:e24}), (\ref{eqn:e30}), and (\ref{eqn:e31}), where 
       the calculation of the inverse of the matrix $S$ is also included in estimating  
       the computational cost of Eq.~(\ref{eqn:e24}), and the sparse structure in
       the matrix $B$ is taken into account for Eqs.~(\ref{eqn:e20}) and (\ref{eqn:e24}). 
       }
   \vspace{1mm}
   \begin{tabular}{lcccc}
   \hline\hline
          & & 1D & 2D & 3D\\
     \hline
     Eq.~(\ref{eqn:e20}) & \quad& $(\log_2N)^2\quad$ & $N^{3/2}\log_2N\quad$ & $N^2\quad$ \\
     Eq.~(\ref{eqn:e21}) & \quad& $N(\log_2N)^2$  & $N^2$ & $N^{7/3}$ \\
     Eq.~(\ref{eqn:e22}) & & $N\log_2N$ &  $N^{3/2}$ & $N^{5/3}$ \\
     Eq.~(\ref{eqn:e24}) & & $\log_2N$ & $N$ & $N^{4/3}$ \\
     Eq.~(\ref{eqn:e30}) & & $N$ & $N^{3/2}$ & $N^{5/3}$ \\
     Eq.~(\ref{eqn:e31}) & & $N\log_2N$ & $N^2$ & $N^{7/3}$ \\
  \hline
   \end{tabular}
  \end{table}

For the finite 3D cubic lattice system we have $N^{(\rm 1)}_{m}=N/(2^{P-m})$ 
as well as the 1D and 2D systems. As shown in the analysis of the 2D systems, 
by approximating $N^{(\rm 2)}_{m}$ and $N^{(\rm 3)}_{p}$ as 
$N^{(\rm 2)}_{m}\approx N^{2/3}/2^{\frac{2}{3}(P-m-1)}$ and 
$N^{(\rm 3)}_{p}\approx N^{2/3}/2^{\frac{2}{3}(P-p)}$, 
which are equal to or more than the corresponding exact number,
we can estimate the computational cost $t_{\rm 3D}$ for the 3D system 
as follows:
\begin{eqnarray}
  \nonumber
  t_{\rm 3D} 
  &\propto& 
  \sum_{p=1}^{P} \sum_{m=0}^{p-1}\sum_{n=0}^{2^{P-m}-1} 
  \frac{N}{2^{P-m}}
   N^{(\rm 2)}_{p,m,n} 
   N^{(\rm 3)}_{p,m,n}\\
  \nonumber
  &<&
  \sum_{p=1}^{P} 
  \sum_{m=0}^{p-1}
  \sum_{n=0}^{2^{P-m}-1}
  \frac{N}{2^{P-m}}
  \frac{N^{2/3}}{2^{\frac{2}{3}(P-m-1)}}
  \frac{N^{2/3}}{2^{\frac{2}{3}(P-p)}}\\
  \nonumber
  &=&
  \frac{4N^{7/3}}{2^{2/3}6-9}
  \left(
    -1+2^{2/3} - \frac{1}{2^{2/3}2^{4P/3}} 
   \right. \\
   &&
   \left.
    + \frac{1}{2^{2/3}2^{2P/3}}
    - \frac{2^{2/3}}{2^{2P/3}}
    + \frac{1}{2^{4P/3}} 
  \right).
\end{eqnarray}
Since we see that the first two terms in parenthesis of the last line 
are the leading term, it is concluded that 
the computational cost for the 3D system scales as O$(N^{7/3})$.

We further analyze the computational cost of the other 
Eqs.~(\ref{eqn:e20}), (\ref{eqn:e22}), (\ref{eqn:e24}),
(\ref{eqn:e30}), and (\ref{eqn:e31}) which are the primary equations 
for the calculation of the Green function. 
Although the detailed derivations are not shown here,
they can be derived in the same way as for Eq.~(\ref{eqn:e21}). 
Table II shows the order of the computational cost for 
each equation. It is found that the computational cost is governed by 
Eq.~(\ref{eqn:e21}), while the computational cost of Eq.~(\ref{eqn:e31}) 
is similar to that of Eq.~(\ref{eqn:e21}).
Thus, it is concluded that as a whole the proposed method scales as 
O$(N(\log_2N)^2)$, O$(N^2)$, and O$(N^{7/3})$ for 1D, 2D, and 3D systems, 
respectively.\cite{comp}

\section{NUMERICAL RESULTS}

In the section several numerical calculations for the $s$-valent NNTB model and DFT
are presented to illustrate the low-order scaling method.
All the DFT calculations in this study were performed by the DFT code OpenMX.\cite{OpenMX} 
The PAO basis functions\cite{PAO} used in the DFT calculations  
are specified by H4.5-$s1$, C5.0-$s1p1$, N4.5-$s1p1$, O4.5-$s1p1$, and P6.0-$s1p1d1$ 
for deoxyribonuleic acid (DNA),
C4.0-$s1p1$ for a single C$_{60}$ molecule, and Pt7.0-$s2p2d1$ for 
a single Pt$_{63}$ cluster, respectively, where the abbreviation of basis function such as 
C5.0-$s1p1$ represents that C stands for the atomic symbol, 5.0 the cutoff radius (bohr)
in the generation by the confinement scheme, $s1p1$ means the employment of one primitive
orbitals for each of $s$ and $p$ orbitals.\cite{PAO} 
Since the PAO basis functions are pseudo-atomic 
orbitals with different cutoff radii depending on atomic species, the resultant Hamiltonian 
and overlap matrices have a disordered sparse structure, reflecting the geometrical structure
of the system.
Norm-conserving pseudopotentials are used in a separable form with multiple projectors 
to replace the deep core potential into a shallow potential.\cite{TM} 
Also a local density approximation (LDA) to the exchange-correlation potential
is employed.\cite{LDA}

\begin{figure}[t]
    \centering
    \includegraphics[width=8.5cm]{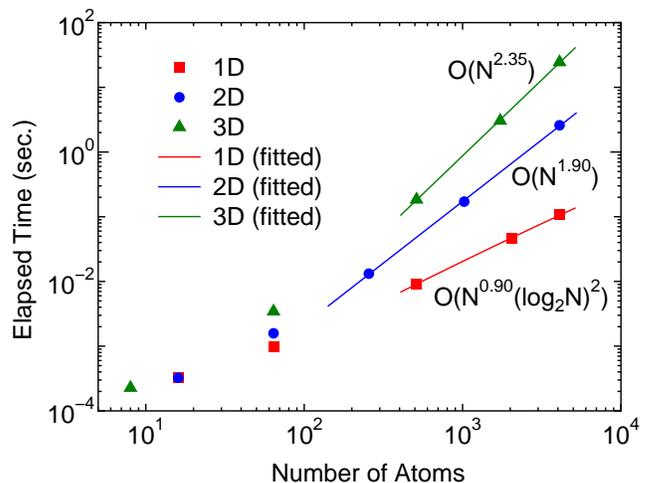}
    \caption{(Color online) 
             The elapsed time of the inverse calculation by Eqs.~(\ref{eqn:e20})-(32)
             for a 1D linear chain, 
             a 2D square lattice, and a 3D cubic lattice systems as a function of number 
             of atoms in the unit cell under periodic boundary condition. 
             The Hamiltonian of the systems are described by 
             the $s$-valent NNTB models. The line for each system is obtained by a least 
             square method, and the computational orders obtained from the fitted curves are 
             O($N^{0.90}(\log_2N)^2$), O($N^{1.90}$), and O($N^{2.35}$) for the 
             1D, 2D, and 3D systems, respectively. 
             The size of domains at the innermost level is set to 20 for all the cases. 
            }
\end{figure}

\subsection{Scaling}

As shown in the previous section, it is possible to reduce the computational cost
from O($N^{3}$) to the low-order scaling without losing numerical accuracy. 
Here we validate the theoretical scaling property of the computational effort by numerical 
calculations. Figure~4 shows the elapsed time required for the calculation of inverse of 
a 1D linear chain, a 2D square lattice, and a 3D cubic lattice systems
as a function of number of atoms in the unit cell under periodic boundary condition, 
which are described by the $s$-valent NNTB models. 
The last three points for each system are fitted to a function by a least square method, 
and the obtained scalings of the elapsed time are found to be 
O($N^{0.90}(\log_2N)^2$), O($N^{1.90}$), and O($N^{2.35}$) 
for the 1D, 2D, and 3D systems, respectively. Thus, we confirm that the scaling of the 
computational cost is nearly the same as that of the theoretical estimation.

\begin{figure}[t]
    \centering
    \includegraphics[width=8.5cm]{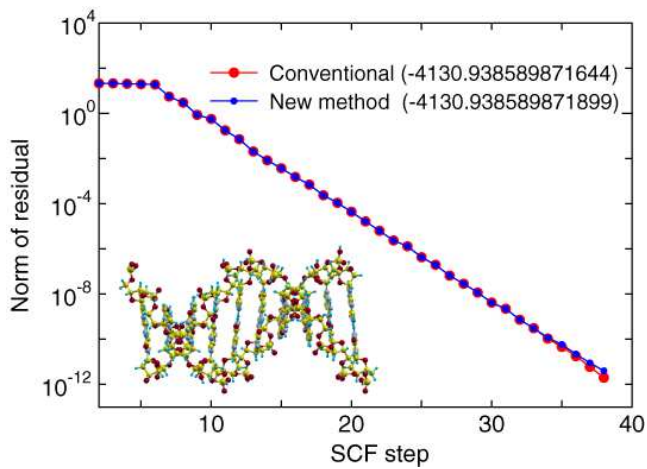}
    \caption{(Color online) 
             The norm of residual in the SCF calculation of DNA, with a periodic 
             double helix structure (650 atoms/unit) consisting of cytosines and guanines,
             calculated by the conventional and proposed methods, 
             where the residual is defined as
             the difference between the input and output charge densities in momentum 
             space. The electric temperature of 700 K and 80 poles for the contour 
             integration are used. 
             The number in parenthesis is the total energy (Hartree) of 
             the system calculated by each method. 
            }
\end{figure}

\subsection{SCF calculation}

To demonstrate that the proposed method is a numerically exact method 
even if the summation in Eq.~(8) is terminated at a modest number of poles,
we show the convergence in the SCF calculations calculated by the conventional 
diagonalization and the proposed methods for deoxyribonuleic acid (DNA)
in Fig.~5, where 80 poles is used for the summation, and the electronic temperature
is 700~K. It is clearly seen that the convergence property and the total energy
are almost equivalent to those by the conventional method with only 80 poles.

\subsection{Iterative search of chemical potential}

Although the computational cost of the proposed method can be reduced from the cubic 
to low-order scalings, the prefactor directly depends on the number of iterations 
in the iterative search of the chemical potential. To address how the combination of 
interpolation and extrapolation methods discussed before works to search a chemical 
potential which conserves the total number of electrons within a criterion, 
we show in Fig.~6 the number of iterations for finding the chemical potential, conserving 
the total number of electrons with a criterion of $10^{-8}$ electron/system, as a function 
of the SCF step for a C$_{60}$ molecule, DNA, and a Pt$_{63}$ cluster. 
Only few iterations are enough to achieve a sufficient convergence of 
the chemical potential as the SCF calculation converges, while a larger number of 
iterations are required at the initial stage of the SCF step. It turns out that 
the proper chemical potential can be searched by the mean iterations of 2.1, 2.4, and 4.0 
for a C$_{60}$ molecule, DNA, and a Pt$_{63}$ cluster, respectively. 
The property of the iterative search is closely related to the energy gap of systems.  
The energy gap between the highest occupied and lowest unoccupied states
of the C$_{60}$ molecule, DNA, and Pt$_{63}$ cluster are 1.95, 0.67, and 0.02~eV, 
respectively.
For the C$_{60}$ molecule and DNA with wide gaps the number of iterations for 
finding the chemical potential tends be large up to 10 SCF iterations, since 
the interpolation or extrapolation scheme may not work well due to the existence of 
the wide gap. 
\begin{figure}[t]
    \centering
    \includegraphics[width=8.5cm]{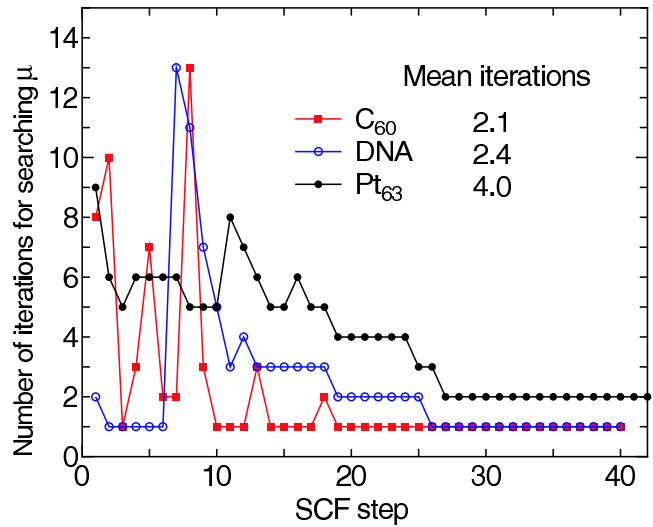}
    \caption{(Color online) 
             The number of iterations for searching the chemical potential 
             which conserves the total number of electrons within a criterion of 
             $10^{-8}$ electron/system for a C$_{60}$ molecule, DNA, and 
             a Pt$_{63}$ cluster, where the electric temperature of 600, 700, 
             and 1000 K, and 80, 80, and 90 poles for the contour integration 
             are used for the C$_{60}$ molecule, DNA, and the Pt$_{63}$ cluster, 
             respectively.
            }
\end{figure}
However, once the charge density nearly converges, the approximate chemical potential in 
between the gap, which is the correct chemical potential at the previous SCF step, 
can satisfy the criterion of $10^{-8}$ electron/system. The situation does correspond
to a small number of iterations after 10 SCF iterations.
Even the trial chemical potential at the first step is the correct one within 
the criterion after 26 SCF iterations in these cases. 
For the Pt$_{63}$ cluster with the narrow gap the number of iterations for finding 
the chemical potential is slightly lower than those of the a C$_{60}$ molecule and DNA
with the wide gaps at the initial stage of SCF iterations, which implies that 
the interpolation and extrapolation schemes by the procedures (3), (4), and (5) can 
give a good estimation of the chemical potential for the nearly continuous eigenvalue 
spectrum. In addition to this, one may find that in contrast to the cases with the wide gap, 
the correct chemical potential is found by two iterations as the charge density converges, 
since a little change of the chemical potential affects the distribution of charge 
density due to the narrow gap. 
However, the fact that only two iterations are sufficient even for the system with 
a narrow gap at the final stage of the SCF step suggests that the extrapolation 
by the procedure (5) works very well. 
Thus, we see from the numerical calculations that the correct chemical potential 
can be searched by only few iterations on an average with the combination of 
interpolation and extrapolation methods for systems with a wide variety of gap.

\begin{figure}[t]
    \centering
    \includegraphics[width=8.5cm]{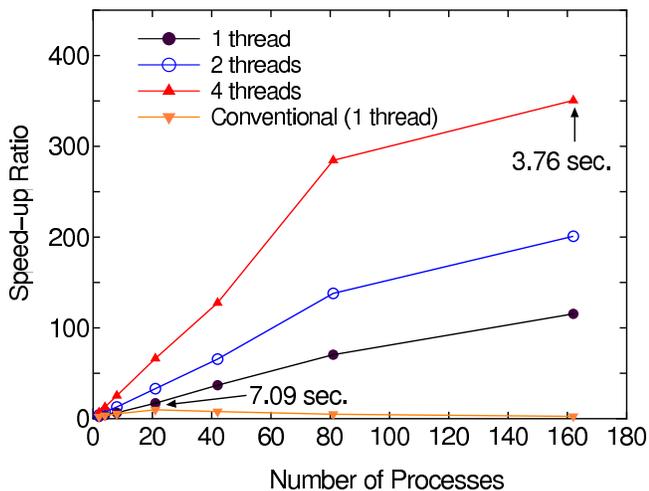}
    \caption{(Color online) 
             Speed-up ratio in the parallel computation of the diagonalization 
             in the SCF calculation for DNA by a hybrid scheme using MPI and OpenMP. 
             The speed-up ratio is defined by $2T_2/T_p$, where $T_2$ and $T_p$ 
             are the elapsed times obtained by two MPI processes and by 
             the corresponding number of processes and threads.
             The structure of DNA is the same as in Fig.~5. 
             The parallel calculations were performed on a Cray XT5 machine consisting 
             of AMD opteron quad core processors (2.3 GHz). 
             The electric temperature of 700 K and 80 poles for the contour 
             integration are used. For comparison, the speed-up ratio for the parallel 
             computation of the conventional scheme using Householder and QR methods is 
             also shown for the case with a single thread.}
\end{figure}

\subsection{Parallel calculation}

We demonstrate that the proposed method is suitable for the parallel computation 
because of the well separated data structure. 
It is apparent that the calculation of the Green function at each $\alpha_p$ in Eq.~(8) 
can be independently performed without data communication among processors. 
Thus, we parallelize the summation in Eq.~(8) by using the message passing interface (MPI) 
in which a nearly same number of poles are distributed to each process. 
The summation in Eq.~(8) can be partly performed in each process, and  
the global summation is completed after all the calculations allocated to each 
process finish. 
In most cases the global summation can be a very small fraction of the computational 
time even including the MPI communication, since the amount of the data to be 
communicated is O($N$) due to the use of localized basis functions. 
In addition to the parallelization of the summation in Eq.~(8), the calculation of 
the Green function can be parallelized in two respects. 
In the recursive calculations of Eqs.~(\ref{eqn:e20})-(32), one may notice that 
the calculation for different $n$ is independently performed, and also 
the calculations involving $V^{T}$ and $L^T$ in Eqs.~(\ref{eqn:e20})-(32) can be 
parallelized with respect to the column of $V^{T}$ and $L^T$ without communication
until the recurrence calculations reach at $m+1=p$. For each $p$ the MPI communication 
only has to be performed at $m+1=p$. In our implementation only the latter part as for 
the calculation of the Green function is parallelized by a hybrid parallelization using MPI 
and OpenMP, which are used for internodes and intranode parallelization. 
As a whole, we parallelize the summation in Eq.~(8) using MPI and the calculations 
involving $V^{T}$ and $L^T$ in Eqs.~(\ref{eqn:e20})-(32) using the hybrid scheme. 

Figure 7 shows the speed-up ratio by the parallel calculation in the elapsed time 
of one SCF iteration. The speed-up ratio reaches about 350 and the elapsed time obtained
is 3.76 $sec$ using 81 processes and 4 threads, which demonstrates the good scalability 
of the proposed method. On the other hand, the conventional diagonalization using 
Householder and QR methods scales up to only 21 processes, which leads to the speed-up 
ratio of 10 and the elapsed time of 7.09 $sec$. Thus, we see that the proposed method 
is of great advantage to the parallel computation unlike the conventional method, while 
the comparison of the elapsed time suggests that the prefactor in the computational effort 
for the proposed method is larger than that of the conventional method.

\section{CONCLUSIONS}

An efficient low-order scaling method has been developed for large-scale DFT calculations 
using localized basis functions such as the PAO, FE, and wavelet basis functions, 
which can be applied to not only insulating but also metallic systems.
The computational effort of the method scales as O($N(\log_2N)^2$), 
O($N^{2}$), and O($N^{7/3}$) for 1D, 2D, and 3D systems, respectively.
The method directly evaluates based on two ideas only selected elements 
in the density matrix which are required for the total energy calculation. 
The first idea is to introduce a contour integration method for the integration of 
the Green function in which the Fermi-Dirac function is expressed by a continued 
fraction. The contour integration enables us to obtain the numerically exact result 
for the integration within double precision at a modest number of poles, which allows
us to regard the method as a numerically exact alternative to conventional 
O($N^{3}$) diagonalization methods. 
It is also shown that the number of poles needed for the convergence does not depend on
the size of the system, but the spectrum radius of the system, which implies that 
the number of poles in the contour integration is unconcerned with 
the scaling property of the computation cost.
The second idea is to employ a set of recurrence formulas for the calculation of 
the Green function. The set of recurrence formulas is derived from a recursive application 
of a block $LDL^T$ factorization using the Schur complement to a structured matrix obtained 
by a nested dissection for the sparse matrix $(ZS-H)$. The primary calculation in the 
recurrence formulas consists of matrix multiplications, and the computational scaling property
is derived by the detailed analysis for the calculations. 
The chemical potential, conserving the total number of electrons, 
is determined by an iterative search which combines several interpolation and 
extrapolation methods. The iterative search permits to find the chemical potential
by less than 5 iterations on an average for systems with a wide variety of gap. 
The good scalability in the parallel computation implies that the method is suitable for 
the massively parallel computation, and could extend the applicability of DFT calculations 
for large-scale systems together with the low-order scaling.

\acknowledgments

The author was partly supported by the Fujitsu lab., the Nissan Motor Co., Ltd., 
Nippon Sheet Glass Co., Ltd., 
and the Next Generation Super Computing Project, Nanoscience Program, 
MEXT, Japan.

\appendix

\section{AN EXAMPLE OF THE INVERSE CALCULATION}

Since the proposed method to calculate the inverse of matrix is largely
difference from conventional methods, we show a simple but nontrivial 
example to illustrate the calculation of the inverse 
by using the set of recurrence formulas 
Eq.~(\ref{eqn:e20})-(\ref{eqn:e24}), (\ref{eqn:e30}), and (\ref{eqn:e31}), 
which may be useful to understand how the calculation proceeds. 
We consider a finite chain molecule consisting of seven atoms described 
by the same $s$-valent NNTB model as in the subsection {\it Nested dissection}, 
where all the on-site energies and hopping integrals are assumed to be 1. 
After performing the nested dissection, we obtain the following structured matrix:
\begin{eqnarray}
  X =  
  \left(
    \begin{array}{ccccccc}
    1 &  & 1 & & & & \\
     & 1 & 1 & & & & 1 \\
    1 & 1 & 1 & & & & \\
    & & & 1 &  & 1 & \\
    & & &  & 1 & 1 & 1\\
    & & & 1 & 1 & 1 & \\
    & 1 & & & 1 &  & 1 \\
    \end{array}
  \right),
\end{eqnarray}
where the blank means a zero element. It can be seen that 
the strucutre is same as in Eq.~(\ref{eqn:e14}), and 
the system is divided into four domains with $P=1$. 
Then, we start from Eqs.~(\ref{eqn:e22}) with $p=0$, 
\begin{eqnarray}
  \nonumber
  V_{0,0,0}^{T} 
  &=& 
  (A_{0,0})^{-1}(B_{0,0}[B_{0,0}])^{T}\\
  &=& 1 \times 1 = 1 \equiv L_{0,0}^{T},\\
  \nonumber
  V_{0,0,1}^{T} 
  &=& 
  (A_{0,1})^{-1}(B_{0,1}[B_{0,1}])^{T}\\
  &=& 1 \times 1 = 1 \equiv L_{0,1}^{T},\\
  \nonumber
  V_{0,0,2}^{T} 
  &=& 
  (A_{0,2})^{-1}(B_{0,2}[B_{0,2}])^{T}\\
  &=& 1 \times 1 = 1 \equiv L_{0,2}^{T},\\
  \nonumber
  V_{0,0,3}^{T} 
  &=& 
  (A_{0,3})^{-1}(B_{0,3}[B_{0,3}])^{T}\\
  &=& 1 \times 1 = 1 \equiv L_{0,3}^{T},
\end{eqnarray}
and proceed to calculate Eq.~(\ref{eqn:e24}),
\begin{eqnarray}
  \nonumber
  S_{0,0} &=& C_{0,0} - B_{0,0}L_{0,0}^{T} - B_{0,1}L_{0,1}^{T}\\
  &=&  1 - 1\times 1- 1\times 1 = -1,\\  
  \nonumber
  S_{0,1} &=& C_{0,1} - B_{0,2}L_{0,2}^{T} - B_{0,3}L_{0,3}^{T}\\
  &=&  1 - 1\times 1- 1\times 1 = -1.  
\end{eqnarray}
$X_{1,0}^{-1}$ and $X_{1,1}^{-1}$ which are precursors of the 
inverse of $X$ can be calculated by Eq.~(\ref{eqn:e30}) 
and (\ref{eqn:e31}) as 
\begin{eqnarray}
  \nonumber
  X_{1,0}^{-1} 
  &=& 
  \left(
    \begin{array}{ccc}
    A^{-1}_{0,0} & * & *\\
    * & A_{0,1}^{-1} & *\\
    * & * & 0
    \end{array}
  \right)\\
 \nonumber
  && + 
  \left(
    \begin{array}{ccc}
    Y_{0,0}^{T}L_{0,0} & * & -Y_{0,0}^{T}\\
    * & Y_{0,1}^{T}L_{0,1} & -Y_{0,1}^{T}\\
    -Y_{0,0} & -Y_{0,1} & S_{0,0}^{-1}
    \end{array}
  \right)\\
 \nonumber 
  &=&
  \left(
    \begin{array}{ccc}
    1 & * & *\\
    * & 1 & *\\
    * & * & 0
    \end{array}
  \right)
   + 
  \left(
    \begin{array}{ccc}
    (-1)\time 1 & *  & -(-1)\\
    * & (-1)\time 1 & -(-1)\\
    -(-1) & -(-1) & -1
    \end{array}
  \right)\\
  &=& 
  \left(
    \begin{array}{ccc}
    0 & * & 1\\
    * & 0 & 1\\
    1 & 1 & -1
    \end{array}
  \right)=X_{1,1}^{-1},
\end{eqnarray}
where $*$ means that the corresponding element is not calculated, and 
remains unknown, since these elements are not referred for further 
calculations.  
The precursor of $X_{1,1}^{-1}$ is found to be same as $X_{1,0}$ due to 
the same inner structure. As the next step, we set $p$ to 1, and calculate 
Eq.~(\ref{eqn:e22}), 
\begin{eqnarray}
  \nonumber
  V_{1,0,0}^{T} 
  &=& 
  (A_{0,0})^{-1}(B_{1,0}[B_{0,0}])^{T}\\
  &=& 1 \times 0 = 0,\\
  \nonumber
  V_{1,0,1}^{T} 
  &=& 
  (A_{0,1})^{-1}(B_{1,0}[B_{0,1}])^{T}\\
  &=& 1 \times 1 = 1,\\
  \nonumber
  V_{1,0,2}^{T} 
  &=& 
  (A_{0,2})^{-1}(B_{1,1}[B_{0,2}])^{T}\\
  &=& 1 \times 1 = 1,\\
  \nonumber
  V_{1,0,3}^{T} 
  &=& 
  (A_{0,3})^{-1}(B_{1,1}[B_{0,3}])^{T}\\
  &=& 1 \times 0 = 0,
\end{eqnarray}
Eq.~(\ref{eqn:e20}),
\begin{eqnarray}
  \nonumber
  Q_{1,1,0}^{T} 
  &=&
  S_{0,0}^{-1}
  \left(
   B_{0,0}V_{1,0,0}^T 
   + 
   B_{0,1}V_{1,0,1}^T 
   -
  (B_{1,0}[C_{0,0}])^{T}
  \right)\\
  &=&
  (-1)(1\times 0 + 1\times 1 - 0) = -1,\\
  \nonumber
  Q_{1,1,1}^{T} 
  &=&
  S_{0,1}^{-1}
  \left(
   B_{0,2}V_{1,0,2}^T 
   + 
   B_{0,3}V_{1,0,3}^T 
   -
  (B_{1,1}[C_{0,1}])^{T}
  \right)\\
  &=&
  (-1)(1\times 1 + 1\times 0 - 0) = -1,
\end{eqnarray}
Eqs.~(\ref{eqn:e21}) and (\ref{eqn:e23}),
\begin{eqnarray}
  \nonumber
  V_{1,1,0}^{T} 
  &=&
  \left(
    \begin{array}{c}
    V_{1,0,0}^{T}\\
    V_{1,0,1}^{T}\\
    0
    \end{array}
  \right)
  + 
  \left(
    \begin{array}{c}
    L_{0,0}^{T}\\
    L_{0,1}^{T}\\
   -1
    \end{array}
  \right)Q_{1,1,0}^{T}\\
  \nonumber
  &=&
  \left(
    \begin{array}{c}
    0\\
    1\\
    0
    \end{array}
  \right)
  + 
  \left(
    \begin{array}{c}
    1\\
    1\\
   -1
    \end{array}
  \right)(-1) 
  = 
  \left(
    \begin{array}{c}
   -1\\
    0\\
    1
    \end{array}
  \right) 
  \equiv L_{1,0}^{T},\\\\
  \nonumber
  V_{1,1,1}^{T} 
  &=&
  \left(
    \begin{array}{c}
    V_{1,0,2}^{T}\\
    V_{1,0,3}^{T}\\
    0
    \end{array}
  \right)
  + 
  \left(
    \begin{array}{c}
    L_{0,2}^{T}\\
    L_{0,3}^{T}\\
   -1
    \end{array}
  \right)Q_{1,1,1}^{T}\\
  \nonumber
  &=&
  \left(
    \begin{array}{c}
    1\\
    0\\
    0
    \end{array}
  \right)
  + 
  \left(
    \begin{array}{c}
    1\\
    1\\
   -1
    \end{array}
  \right)(-1) 
  = 
  \left(
    \begin{array}{c}
    0\\
   -1\\
    1
    \end{array}
  \right) 
  \equiv L_{1,1}^{T},\\
\end{eqnarray}
and Eq.~(\ref{eqn:e24}),
\begin{eqnarray}
 \nonumber
 S_{1,0} 
  &=& 
 C_{1,0} - B_{1,0}L_{1,0}^{T} - B_{1,1}L_{1,1}^{T}\\
 \nonumber
  &=&
 1 - 
  \left(
    \begin{array}{ccc}
    0 & 1 & 0\\
    \end{array}
  \right) 
  \left(
    \begin{array}{c}
   -1\\
    0\\
    1
    \end{array}
  \right)\\
  && 
  -
  \left(
    \begin{array}{ccc}
    1 & 0 & 0\\
    \end{array}
  \right) 
  \left(
    \begin{array}{c}
    0\\
   -1\\
    1
    \end{array}
  \right) = 1. 
\end{eqnarray}
Finally updating the precursors $X_{1,0}^{-1}$ and $X_{1,1}^{-1}$ of the inverse 
of the matrix $X$ using Eqs.~(\ref{eqn:e30}) and (\ref{eqn:e31}) yields 
the inverse of $X$ as follows:
\begin{eqnarray}
  \nonumber
  X_{2,0}^{-1} 
  &=& 
  \left(
    \begin{array}{ccc}
    X_{1,0}^{-1} & * & *\\
    * & X_{1,1}^{-1} & *\\
    * & * & 0
    \end{array}
  \right)\\
  \nonumber
  &&
   + 
  \left(
    \begin{array}{ccc}
    Y_{1,0}^{T}L_{1,0} & * & -Y_{1,0}^{T}\\
    * & Y_{1,1}^{T}L_{1,1} & -Y_{1,1}^{T}\\
    -Y_{1,0} & -Y_{1,1} & S_{1,0}^{-1}
    \end{array}
  \right)\\
  \nonumber 
  &=&
  \left(
    \begin{array}{ccccccc}
    0 & * & 1 & * & * & * & *\\
    * & 0 & 1 & * & * & * & 0\\
    1 & 1 &-1 & * & * & * & *\\
    * & * & * & 0 & * & 1 & 0\\
    * & * & * & * & 0 & 1 & *\\
    * & * & * & 1 & 1 &-1 & *\\
    * & 0 & * & 0 & * & * & 0\\
    \end{array}
  \right)\\
  \nonumber
  &&
   + 
  \left(
    \begin{array}{ccccccc}
    1 & * &-1 & * & * & * & *\\
    * & 0 & 0 & * & * & * & 0\\
   -1 & 0 & 1 & * & * & * & *\\
    * & * & * & 0 & * & 0 & 0\\
    * & * & * & * & 1 &-1 & *\\
    * & * & * & 0 &-1 & 1 & *\\
    * & 0 & * & 0 & * & * & 1\\
    \end{array}
  \right)\\
  &=& 
  \left(
    \begin{array}{ccccccc}
    1 & * & 0 & * & * & * & *\\
    * & 0 & 1 & * & * & * & 0\\
    0 & 1 & 0 & * & * & * & *\\
    * & * & * & 0 & * & 1 & 0\\
    * & * & * & * & 1 & 0 & *\\
    * & * & * & 1 & 0 & 0 & *\\
    * & 0 & * & 0 & * & * & 1\\
    \end{array}
  \right) \equiv X^{-1}.
\end{eqnarray}
The calculated elements in the inverse $X^{-1}$ are found to be consistent 
with those by conventional methods such as the LU method. It is also noted that
one can easily obtain the corresponding elements in the inverse of the 
original matrix using a table function generated in the nested dissection 
which converts the row or column index of the structured matrix to the 
original one.

\section{Calculation of selected eigenstates}

In the appendix, it is shown that 
a few eigenstates around a selected energy $\xi$ can be obtained by a similar 
way with the same computational complexity as in the calculation 
for the density matrix, though the proposed method directly computes 
the density matrix without explicitly calculating the eigenvectors. 

We compute the few eigenstates around $\xi$ using a block shift-invert iterative 
method in which the generalized eigenvalue problem of Eq.~(2) is transformed as 
\begin{eqnarray}
  (H-\xi S)^{-1}Sc_{\nu} = \frac{1}{\varepsilon_{\nu}-\xi}c_{\nu}.  
\end{eqnarray}
Then, the following iterative procedure yields a set of eigenstates around 
$\xi$ as the convergent result.
\begin{eqnarray}
  {\bf b}_{l} &=& (H-\xi S)^{-1}S{\bf c}_{l},
\end{eqnarray}
\begin{eqnarray}
  \langle {\bf b}_{l} 
  \vert \hat{H} \vert {\bf b}_{l} \rangle
   {\bf c}_{l+1}
   &=& 
  \langle {\bf b}_{l} 
  \vert \hat{S} \vert {\bf b}_{l} \rangle
  {\bf c}_{l+1}
  \underline{\varepsilon}_{l+1},
\end{eqnarray}
where $l$ is the iterative step, $\underline{\varepsilon}$ is a square 
matrix consisting of diagonal elements, and ${\bf b}$ and ${\bf c}$ 
are a set of vectors of which number is that of the selected states. 
The matrix multiplication in Eq.~(B2) and the solution of 
the generalized eigenvalue problem for Eq.~(B3) are repeated until 
convergence, and the convergent ${\bf c}$ and the diagonal elements 
of $\underline{\varepsilon}$ correspond to the eigenstates around $\xi$.
If the number of selected eigenstates is independent of the size of system, 
the computational cost required for Eq.~(B3) is O($N$), which arises from 
the matrix multiplications of 
$\langle {\bf b}_{l} 
\vert \hat{H} \vert {\bf b}_{l} \rangle
$
and 
$\langle {\bf b}_{l} 
\vert \hat{S} \vert {\bf b}_{l} \rangle
$.
Therefore, the computational cost of the iterative calculation is governed 
by the matrix multiplication of $(H-\xi S)^{-1}{\bf y}^{T}_{l}$ in Eq.~(B4), 
where ${\bf y}^{T}_{l} =S{\bf c}_{l}$. 

Here we show that the matrix multiplication of $(H-\xi S)^{-1}{\bf y}^{T}_{l}$ 
can be performed by a similar way with the same computational complexity 
as in the calculation for the density matrix. 
As an example of $(H-\xi S)$, let us consider the matrix $X$ given by 
Eq.~(\ref{eqn:e14}). After the recurrence calculation of 
Eqs.~(\ref{eqn:e20})-(32), it turns out that the matrix $X$ is factorized as
\begin{eqnarray}
  X = L_{1}L_{0} D L_{0}^{T}L_{1}^{T}  
\end{eqnarray}
with matrices defined by 
\begin{eqnarray} 
  \nonumber
  D =  
  {\footnotesize
  \left(
    \begin{array}{ccccccc}
    A_{0,0} &  & & & & & \\
    & A_{0,1} & & & & & \\
    & & S_{0,0} & & & & \\
    & & & A_{0,2} & & & \\
    & & & & A_{0,3} & & \\
    & & & & & S_{0,1} & \\
    & & & & & & S_{1,0} \\
    \end{array}
  \right)},
\end{eqnarray}
\begin{eqnarray}
  \nonumber
  L_{0} = 
  {\footnotesize
  \left(
    \begin{array}{ccccccc}
    I_{A_{0,0}} &  & & & & & \\
     & I_{A_{0,1}} & & & & & \\
    L_{0,0} & L_{0,1} & I_{C_{0,0}} & & & & \\
    & & & I_{A_{0,2}} &  & & \\
    & & &  & I_{A_{0,3}} & & \\
    & & & L_{0,2} & L_{0,3} & I_{C_{0,1}} & \\
    & & & & &  & I_{C_{1,0}} \\
    \end{array}
  \right)},
\end{eqnarray}
and 
\begin{eqnarray}
  \nonumber
  L_{1} = 
  {\footnotesize
  \left(
    \begin{array}{ccccccc}
    I_{A_{0,0}} &  & & & & & \\
    & I_{A_{0,1}} & & & & & \\
    & & I_{C_{0,0}} & & & & \\
    & & & I_{A_{0,2}} &  & & \\
    & & &  & I_{A_{0,3}} & & \\
    & & & & & I_{C_{0,1}} & \\
    & L_{1,0} & & & L_{1,1} &  & I_{C_{1,0}} \\
    \end{array}
  \right)},
\end{eqnarray}
where $I_{A_{0,0}}$ stands for an identity matrix with the same size as 
that of the matrix $A_{0,0}$, and the same rule applies to other cases. 
Then, we see that the inverse of $X$ is given by 
\begin{eqnarray}
  X^{-1} = (L_{1}^{T})^{-1}(L_{0}^{T})^{-1} D^{-1} (L_{0})^{-1}(L_{1})^{-1}
\end{eqnarray}
with matrices defined by 
\begin{eqnarray}
  \nonumber
  (L_{0})^{-1} =  
  {\footnotesize
  \left(
    \begin{array}{ccccccc}
    I_{A_{0,0}} &  & & & & & \\
     & I_{A_{0,1}} & & & & & \\
   -L_{0,0} &-L_{0,1} & I_{C_{0,0}} & & & & \\
    & & & I_{A_{0,2}} &  & & \\
    & & &  & I_{A_{0,3}} & & \\
    & & &-L_{0,2} &-L_{0,3} & I_{C_{0,1}} & \\
    & & & & &  & I_{C_{1,0}} \\
    \end{array}
  \right)},
\end{eqnarray}
and 
\begin{eqnarray}
  \nonumber
  (L_{1})^{-1} =  
  {\footnotesize
  \left(
    \begin{array}{ccccccc}
    I_{A_{0,0}} &  & & & & & \\
    & I_{A_{0,1}} & & & & & \\
    & & I_{C_{0,0}} & & & & \\
    & & & I_{A_{0,2}} &  & & \\
    & & &  & I_{A_{0,3}} & & \\
    & & & & & I_{C_{0,1}} & \\
    &-L_{1,0} & & &-L_{1,1} &  & I_{C_{1,0}} \\
    \end{array}
  \right)}.
\end{eqnarray}
It should be noted that the inverses of $L_0$ and $L_1$ are remarkably simple, 
and that the inverse of $D$ is found to be a matrix consisting of diagonal block 
inverses. 
In general cases, we see that a matrix $X$ and its inverse are given by 
\begin{eqnarray}
  X = L_{P}\cdots L_{1}L_{0} D L_{0}^{T}L_{1}^{T}\cdots L_{P}^{T},
\end{eqnarray}
\begin{eqnarray}
  \nonumber
  X^{-1} &=& (L_{P}^{T})^{-1}\cdots (L_{1}^{T})^{-1}(L_{0}^{T})^{-1} 
              D^{-1}\\
      &&   \times(L_{0})^{-1}(L_{1})^{-1}\cdots (L_{P})^{-1},
\end{eqnarray}
where the inverse $L_p$ is given in a similar form as well as those 
of $L_0$ and $L_1$.

By considering Eq.~(B7) and the simple forms of $(L_p)^{-1}$, 
the matrix multiplication of $X^{-1}{\bf y}^{T}$ can be performed by 
the following three steps:\\

\noindent
(i) {\it The first step},  
$({\bf y}')^{T}=(L_{0})^{-1}(L_{1})^{-1}\cdots (L_{P})^{-1}{\bf y}^{T}$, 
is calculated by  
\begin{eqnarray}
  \nonumber
   \label{eqn:b8} 
  ({\bf y}'[I_{C_{p,n}}])^{T} &=& -L_{p,2n}({\bf y}[L_{p,2n}])^{T}
                    -L_{p,2n+1}({\bf y}[L_{p,2n+1})^{T}\\
                 && +({\bf y}[I_{C_{p,n}}])^{T},\\
   \label{eqn:b9}
  ({\bf y}'[I_{A_{0,n}}])^{T} &=& ({\bf y}[I_{A_{0,n}}])^{T},
\end{eqnarray}
where $p=0,\cdots,P$ and $n=0,\cdots,2^{P-p}-1$ in Eq.~(B8), 
and $n=0,\cdots,2^{P}-1$ in Eq.~(B9).\\ 

\noindent
(ii) {\it The second step},  
$({\bf y}'')^{T}=D^{-1}({\bf y}')^{T}$, 
is calculated by  
\begin{eqnarray}
   \label{eqn:b10}
  ({\bf y}''[I_{C_{p,n}}])^{T} &=& (S_{p,n})^{-1}({\bf y}'[S_{p,n}])^{T},\\
   \label{eqn:b11}
  ({\bf y}''[I_{A_{0,n}}])^{T} &=& (A_{0,n})^{-1}({\bf y}'[I_{A_{0,n}}])^{T},
\end{eqnarray}
where $p=0,\cdots,P$ and $n=0,\cdots,2^{P-p}-1$ in Eq.~(B10), 
and $n=0,\cdots,2^{P}-1$ in Eq.~(B11).\\ 

\noindent
(iii) {\it The third step},  
$(L_{P}^{T})^{-1}\cdots (L_{1}^{T})^{-1}(L_{0}^{T})^{-1} ({\bf y}'')^{T}$, 
is performed by the following recurrence formulas:
\begin{eqnarray}
  \nonumber
   \label{eqn:b12}
  ({\bf x}_{p+1}[L_{p,2n}])^{T} &=& 
    ({\bf x}_{p}[L_{p,2n}])^{T} - (L_{p,2n})^{T}({\bf x}_{p}[I_{C_{p,n}}])^{T},\\\\
  \nonumber
  ({\bf x}_{p+1}[L_{p,2n+1}])^{T} &=& 
    ({\bf x}_{p}[L_{p,2n+1}])^{T}\\
    &&
    -(L_{p,2n+1})^{T}({\bf x}_{p}[I_{C_{p,n}}])^{T},\label{eqn:b13}\\
  ({\bf x}_{p+1}[I_{C_{m,n}}])^{T} &=& ({\bf x}_{p}[I_{C_{m,n}}])^{T},\label{eqn:b14}
\end{eqnarray}
where ${\bf x}_{0} = {\bf y}''$, $p+1=1, \cdots, P+1$, and $m=p,\cdots, P$.
At the end of the recurrence calculation, we obtain the result of 
the multiplication as
\begin{eqnarray}
   \label{eqn:b15}
   X^{-1}{\bf y}^{T} = {\bf x}_{P+1} \equiv (H-\xi S)^{-1}{\bf y}^{T}.
\end{eqnarray}

The computational effort of the three steps can be easily estimated by the same way 
as for the calculation of the inverse matrix, and summarized in Table III. It is found 
that the the computational complexity of the three steps is lower than that of the 
calculation of the inverse matrix. Thus, if the number of selected eigenstates 
and the number of iterations for convergence are independent of the size of system, 
the computational effort of calculation of the selected eigenstates is governed by 
the recurrence calculation of Eqs.~(\ref{eqn:e20})-(\ref{eqn:e24}) even for 
the calculation of selected eigenstates. The scheme may be useful for calculation of 
eigenstates near the Fermi level.

   \begin{table}[t]
     \caption{
       Computational order of Eqs.~(\ref{eqn:b8}), (\ref{eqn:b10}), (\ref{eqn:b11}),
       (\ref{eqn:b12}), and (\ref{eqn:b13}).
       }
   \vspace{1mm}
   \begin{tabular}{lcccc}
   \hline\hline
          & & 1D & 2D & 3D\\
     \hline
     Eq.~(\ref{eqn:b8})  & & $N\log_2N\quad$ & $N^{3/2}\quad$ & $N^{5/3}\quad$ \\
     Eq.~(\ref{eqn:b10}) & & $N$  & $N\log_2N$ & $N^{4/3}$ \\
     Eq.~(\ref{eqn:b11}) & & $N$ &  $N$ & $N$ \\
     Eqs.~(\ref{eqn:b12})
        +(\ref{eqn:b13}) & & $N\log_2N$ & $N^{3/2}$ & $N^{5/3}$ \\
  \hline
   \end{tabular}
  \end{table}


\begin{thebibliography}{99}
\bibitem{Muller_add}
Although the coefficients $a$, $b$, and $c$ can be analytically evaluated,
the round-off error in the analytically evaluated solution is nonnegligible
as $\Delta N$ converges to zero. In order to avoid the numerical problem, 
we refine $a$, $b$, and $c$ by minimizing a function 
$F\equiv\sum_{i=0}^{2}(a\mu_i^2 + b\mu_i + c - \Delta N_i)^2$ with 
the analytically evaluated coefficients as initial values.
We find that the coefficients refined by the minimization are much 
more accurate than the analytic ones for serious cases, and that the 
refinement is quite effective to avoid the numerical instability.
\end{thebibliography}
\end{document}